\documentclass[twocolumn,linenumbers]{aastex631}
\usepackage{amssymb,amsmath}
\usepackage{verbatim}
\allowdisplaybreaks[1]
\shorttitle{Dust maps for Cosmology}
\shortauthors{Mudur et al.}
\graphicspath{{./}{figure*s/}}

\begin{document}
\nolinenumbers

\title{Stellar Reddening Based Extinction Maps for Cosmological Applications}

\author[0000-0001-5139-612X]{Nayantara Mudur}
\affiliation{Department of Physics, Harvard University, 17 Oxford St., Cambridge, MA 02138, USA}
\affiliation{Harvard-Smithsonian Center for Astrophysics, 60 Garden St., Cambridge, MA 02138, USA}

\author[0000-0002-9542-2913]{Core Francisco Park}
\affiliation{Department of Physics, Harvard University, 17 Oxford St., Cambridge, MA 02138, USA}

\author[0000-0003-2808-275X]{Douglas P. Finkbeiner}
\affiliation{Department of Physics, Harvard University, 17 Oxford St., Cambridge, MA 02138, USA}
\affiliation{Harvard-Smithsonian Center for Astrophysics, 60 Garden St., Cambridge, MA 02138, USA}



\newcommand\change[1]{{\color{orange}#1}}

\begin{abstract}
Cosmological surveys must correct their observations for the reddening of extragalactic objects by Galactic dust. Existing dust maps, however, have been found to have spatial correlations with the large-scale structure of the Universe. Errors in extinction maps can propagate systematic biases into samples of dereddened extragalactic objects and into cosmological measurements such as correlation functions between foreground lenses and background objects and the primordial non-gaussianity parameter $f_{NL}$. Emission-based maps are contaminated by the cosmic infrared background, while maps inferred from stellar-reddenings suffer from imperfect removal of quasars and galaxies from stellar catalogs. Thus, stellar-reddening based maps using catalogs without extragalactic objects offer a promising path to making dust maps with minimal correlations with large-scale structure. We present two high-latitude integrated extinction maps based on stellar reddenings, with a point spread function of full-width half-maximum 6.1' and 15'. We employ a strict selection of catalog objects to filter out galaxies and quasars and measure the spatial correlation of our extinction maps with extragalactic structure. Our galactic extinction maps have reduced spatial correlation with large scale structure relative to most existing stellar-reddening based and emission-based extinction maps.

\end{abstract}


\newcommand\citehere[1]{{\color{blue} add citation! #1}}
\newcommand\todo[1]{{\color{blue}#1}}
\newcommand\ns[1]{\textsc{Nside}=#1}
\newcommand\bay[1]{Bayestar#1}
\newcommand\com[1]{
}

\section{Introduction} \label{sec:intro}
Upcoming Stage IV cosmological surveys, such as the Dark Energy Spectroscopic Instrument (DESI) \citep{collaboration2016desi} and the  Vera C. Rubin Observatory Legacy Survey of Space and Time (LSST) \citep{ivezic2019lsst} will image tens of millions of galaxies, supernovae and quasi-stellar objects, and aim to constrain cosmological parameters at the sub-percent level. To maximally leverage the statistical power of these datasets, it is imperative to characterize and minimize possible sources of systematic errors, such as Galactic dust extinction.

The interstellar medium of the Milky Way is speckled with dust. Interstellar dust is valuable as a probe of the interstellar medium and in studies of molecular clouds and structures in the Galaxy. However, it also obscures and modifies our view of both stars and extragalactic objects due to absorption and scattering of light by the intervening dust column, an effect referred to as extinction. Thus, dust maps are needed to subtract the contribution of dust extinction from the observed magnitude of a background object prior to performing subsequent cosmological analyses. 

Dust maps can be broadly categorized into two classes -- emission-based maps and extinction-based maps. Emission-based maps use thermal emission at far infrared wavelengths to derive the dust column density. The Schlegel-Finkbeiner-Davis map \citep[][hereafter SFD]{Schlegel1997} combined the high angular resolution of the Infrared Astronomy Satellite \citep[IRAS;][]{beichman1988infrared} map at 100$\mu$m with the Diffuse Infrared Background Experiment \citep[DIRBE;][]{hauser1998cobe} maps' superior calibration and wavelength coverage to derive a dust map with an angular resolution of full-width half maximum = $6.1'$. SFD, along with the Planck thermal dust maps \citep{aghanim2016planck} at 353, 545 and 857 GHz and the Planck dust optical depth map, have since been among the most common choices of maps for applications in cosmology. However, \cite{yahata2007effect} found that over regions of the sky with an average extinction of $A_{r, SFD}<0.1$~mag, the number counts of galaxies increased and their average post-extinction correction color was bluer with increasing extinction -- contrary to the effect physically expected due to the obscuration of dust. They posited that such an effect could be explained by localized fluctuations due to far infrared emission from background galaxies in the SFD extinction map. Furthermore, \cite{chiang2019extragalactic} found that out of ten emission and extinction-based maps, all except the H\textsc{i}-based \cite{lenz2017new} map possess significant correlations with extragalactic structure. In emission-based maps, this arises from imperfect removal of the cosmic infrared background, the cumulative emission of dusty star-forming galaxies \citep{lagache2005dusty,dole2006cosmic}. 

Another dust mapping approach involves using inferred stellar extinctions, conditional on an inferred or fixed reddening law. The line-of-sight reddening to a star can be derived by combining theoretical knowledge of stellar types and intrinsic magnitudes with observed photometry from surveys. Since each star traces the extinction up to its distance, stellar reddening based maps allow dust extinction to be reconstructed in three dimensions. For these maps, contamination arises from imperfect removal of galaxies and quasi-stellar objects (QSOs) from stellar catalogs, leading to systematic overestimation and underestimation of extinction at their positions. 

Correlations of galactic extinction maps with large-scale structure are concerning because when dust maps are used to deredden extragalactic objects, they imprint biases on the objects' dereddened magnitudes, as in \cite{yahata2007effect}, which can further affect downstream analyses. \cite{chiang2019extragalactic} highlight secondary effects such as a bias on cosmological parameter constraints derived from luminosity distance measurements of dereddened Type Ia Supernovae, and on overdensities and two-point correlation functions estimated from magnitude-limited samples of extragalactic objects that can further propagate into correlations between foreground lenses and background objects. These lensing-induced correlation functions \citep[e.g.,][]{menard2002cosmological, garcia2018weak} are important probes of the galaxy bias, or the relation between foreground galaxy overdensities and matter overdensities \citep{dodelson2016cosmic}. Other recent work \citep{awan2016testing, kitanidis2020imaging} has also shown that dust extinction can be a significant source of systematics in other cosmological measurements. Since stellar-extinction based maps should only be contaminated by the presence of stray extragalactic objects in stellar catalogs, in principle, maps built using stellar catalogs that have these stray galaxies and QSOs removed, should offer a promising path to generating maps without correlations with large-scale structure.

In this paper, we describe a set of two stellar reddening-based maps with a well-defined point-spread function. We take more care to eliminate galaxies and QSOs from our star sample, and analyze the extent of correlation of the resulting maps with extragalactic structure. The document is organized as follows: Section \ref{sec:datasets} describes the datasets used in the inference and analysis process, Section \ref{sec:selections} describes the stellar selections we use to eliminate galaxies and quasars. Section \ref{sec: reconstruction} describes the steps used in our reconstruction. Section \ref{sec:analysis} examines the performance of the stellar selections on a spectroscopically-matched test sample, measures the spatial correlation with large scale structure of the maps we make in addition to other dust maps in the literature and assesses the extent of noise in different parts of the sky. Section \ref{sec:discussion} discusses potential uses of the maps and future work.


\section{Datasets} \label{sec:datasets}
In this section we describe three datasets used in this work: 1.\ the star sample, including data used to remove galaxies and QSOs; 2.\ the spectroscopically matched sample used to tune those cuts; and 3.\ the extragalactic objects used as tracers of large-scale structure in Section~\ref{sec:analysis}.

\subsection{The Star Sample}
Stellar-extinction maps are built using inferred per-star posteriors on reddening and distance modulus for objects from stellar catalogs. For each star, we constrain the posterior using the Bayestar stellar inference framework \citep{Green2019}, inputting magnitudes from Pan-STARRS1 and 2MASS along with Gaia parallaxes.

The Panoramic Survey Telescope and Rapid Response System (Pan-STARRS1) is a ground-based sky survey using the eponymous 1.8-meter diameter telescope on the island of Maui, in Hawaii \citep{PS1_paper1, PS1_paper6}. The first data release, the $3\pi$ Steradian Survey, imaged three quarters of the sky north of $\delta=-30^{\circ}$ in 5 broadband filters denoted by $g_{P1}, r_{P1}, i_{P1}, z_{P1}, y_{P1}$ -- with mean wavelengths of 490nm, 624nm, 756nm, 869nm and 964nm -- from May 2010 through March 2014. Sources are detected and processed by the Pan-STARRS Image Processing Pipeline \textsc{psphot} to generate photometric and astrometric measurements \citep{PS1_paper2, PS1_paper4, PS1_paper3}. Photometric calibration was performed using the `ubercalibration' approach \citep{padmanabhan2008improved, finkbeiner2016hypercalibration}, that solves for photometric parameters by minimizing the variance of the flux from repeated measurements of the same stars \citep{schlafly2012photometric, PS1_paper5}, resulting in a precision of around $10$~mmag in all 5 bands.

The Two Micron All Sky Survey \citep[2MASS;][]{cutri20032mass, skrutskie2006two} was a ground-based full-sky survey that scanned the sky in three bands in the near-infrared -- J, H, and K$_s$, centered at $1.25\mu$m, $1.65\mu$m and $2.16\mu$m -- from June 1997 to February 2001 and produced a catalog of 471 million sources. The survey used two 1.3m telescopes, one at Mt. Hopkins, Arizona, and the other at the Cerro Tololo Inter-American Observatory (CTIO) in Chile. 

The \textit{Gaia} mission \citep{gaia2016gaia, prusti2016gaia} is a space-based mission that measures the three dimensional spatial and kinematic distribution of stars in the Galaxy. The satellite consists of three instruments -- the astrometric instrument, that collects images in the broad G band (330-1050 nm), the blue (BP, 330-680nm) and red (RP, 630-1050nm) prism photometers and a radial velocity spectrometer. The astrometric instrument on the satellite consists of two identical three-mirror anastigmat (TMA) telescopes with apertures of  1.45m$\times$0.50m  with a shared focal plane, and uses the principle of scanning space astrometry \citep{turon2010basic}. The mission has measured the parallaxes and proper motions of over a billion stars, or around 1\% of all the stars in the Milky Way.  
The Gaia Early Data Release 3 \citep[][hereafter Gaia EDR3]{brown2021gaia} consists of data collected in the first 34 months of the mission and includes the proper motions and parallaxes of over 1.8 billion sources brighter than G=21.

We use additional photometric bands from SDSS and WISE for each candidate star to reject galaxies and QSOs. These fields are not used in deriving the posteriors on distance modulus and reddening, or in our preliminary selections, but are used in the secondary selections (Section \ref{subsec:cuts}) to minimize contamination from extragalactic objects.

The Wide-field Infrared Survey Explorer \citep[WISE;][]{wright2010wide} mapped the sky in four infrared bands, centered at 3.4, 4.6, 12, and 22 $\micron$ -- referred to as the W1, W2, W3, and W4 bands. The survey used a 40 cm cryogenically-cooled telescope. The 4-Band Cryogenic survey lasted from January 7 2010 to August 6 2010, followed by a 3-Band Cryo phase and the NEOWISE Post-Cryo mission \citep{mainzer2011preliminary} phase that ended on February 1, 2011. The AllWISE Data Release \citep{cutri2013explanatory}, that included the AllWISE Source Catalog, combined data from all three phases. The AllWISE Source Catalog consisted of the astrometry and photometry of over 747 million objects, with typically higher sensitivity in the W1 and W2 bands than previous WISE releases.

The Sloan Digital Sky Survey \citep[SDSS;][]{blanton2017sloan} is an imaging and spectroscopic survey that started in 1998, and uses the 2.5 m Sloan Foundation Telescope at the Apache Point Observatory, in New Mexico \citep{gunn20062}. SDSS measured photometry in the $u$, $g$, $r$, $i$ and $z$ bands as well as the spectra of millions of stars, galaxies and quasars. We query SDSS Data Release 14 \citep{abolfathi2018fourteenth} $ugriz$ photometry for objects in our sample. We use the Large Survey Database framework \citep{juric2012lsd} to cross match between datasets.

\subsection{The Spectroscopically Matched Subset}
To train the color cuts described in Section \ref{subsec:cuts} and the Appendix and to examine the efficacy of our secondary selections on a subset of labeled objects, we take a sample of stars from the catalog obtained after the preliminary set of selections for $b\rangle50^{\circ}$ (Section 3.1) and cross match to the SDSS Data Release 17 `SpecObj' catalog \citep{accetta2022seventeenth}, requiring a detection in the latter. This catalog consists of all objects whose spectra were obtained, and includes the redshift estimated from the spectra as well as their spectroscopic class.  The spectra of objects usually allow us to reliably identify them as quasars, galaxies or stars, since each of these classes typically have characteristic features that distinguish them. Thus this subset offers a testbed with true labels that allows us to examine how well a given set of stellar selections removes galaxies and quasars.

\subsection{Extragalactic Reference Objects}
\begin{figure}[!h]
\centering
    \includegraphics[keepaspectratio=true, width=0.85\linewidth,height=200pt]{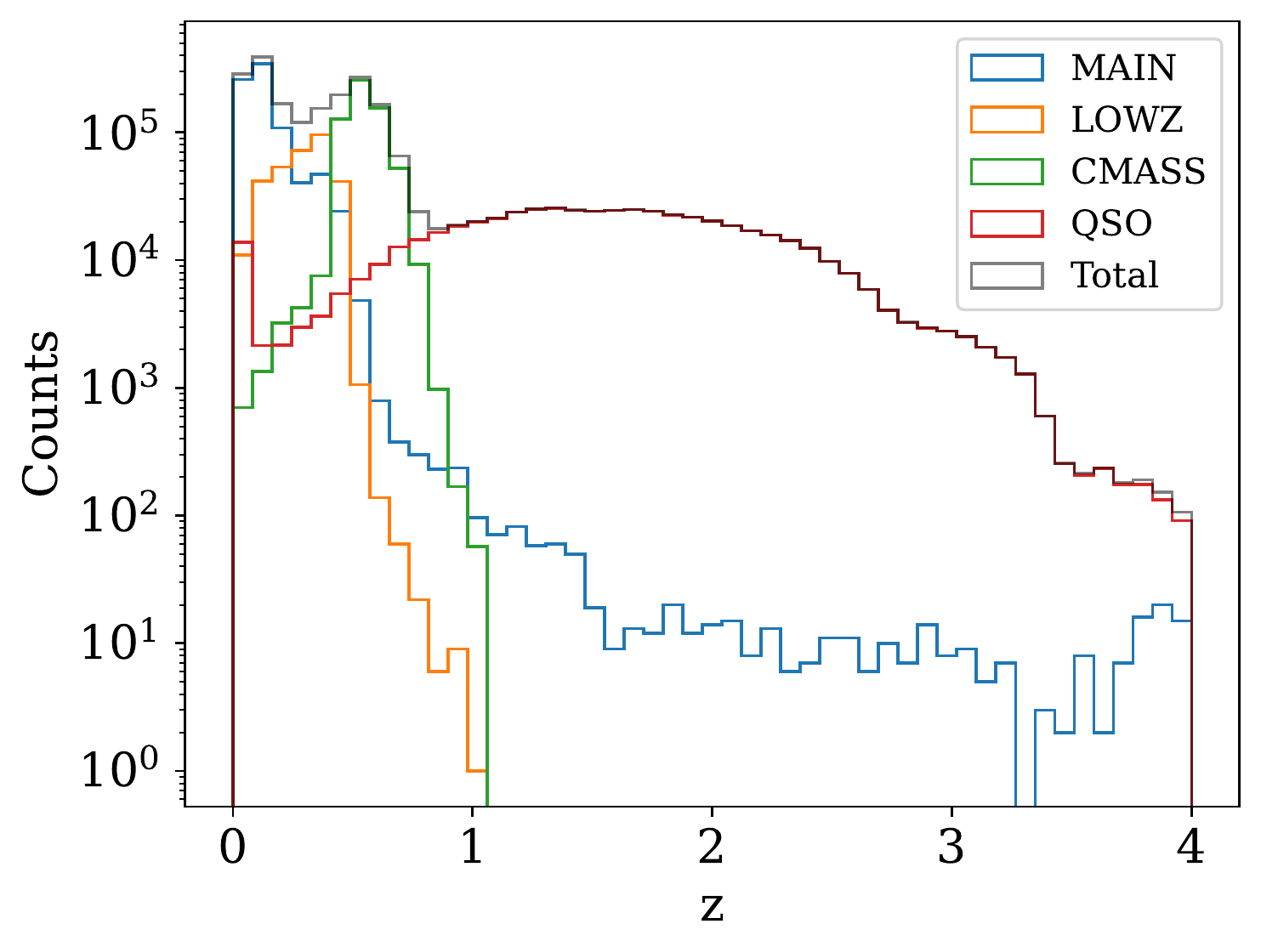}
\caption{Redshift distribution of extragalactic reference objects in the Northern Galactic Hemisphere. These objects were used to compute the angular cross-correlation with LSS.
\label{fig:exgal_dbn}}
\end{figure}

Following \cite{chiang2019extragalactic}, we use four spectroscopic reference samples as tracers of the LSS: the Main galaxy sample from the New York University Value-Added Galaxy Catalog \citep{blanton2005new, adelman2007fifth, padmanabhan2008improved}, BOSS LOWZ and CMASS LRGs \citep{reid2016sdss} and the Data Release 14 Quasar catalog (DR14Q) from the extended Baryon Oscillation Spectroscopic Survey (SDSS-IV eBOSS) \citep{paris2018sloan}. The Main Galaxy sample is similar to the SDSS Main galaxy sample described by \cite{strauss2002spectroscopic} but slightly more inclusive owing to a fainter magnitude limit (18 vs 17.77) and the absence of the exclusion of small bright objects, among other changes. This sample mostly spans $z\lesssim0.3$. The LOWZ and CMASS samples from the Baryon Oscillation Spectroscopic Survey (BOSS) span $z\lesssim0.4$ and $0.4 \lesssim z \lesssim 0.7$. The DR14Q catalog compiled all spectroscopically-confirmed quasars from SDSS I through IV in the redshift range $0 \lesssim z \lesssim 5$. The distribution of this set of extragalactic reference objects is plotted in Figure \ref{fig:exgal_dbn}, similar to Figure 5 in \cite{chiang2019extragalactic}.

\section{Stellar Posterior Inference and Selections} \label{sec:selections}
\subsection{Preliminary Selections} \label{sec:prelim-selections}
Objects are crossmatched from the Pan-STARRS1 catalog to the 2MASS catalog, and the Gaia EDR3 catalog, requiring a detection in Gaia EDR3. We additionally crossmatch to the AllWISE, and the SDSSDR14 datasets. Note, all matches other than the one with Gaia are not exclusive and do not require a detection in the target survey. All survey crossmatches use a radius of 1 arcsecond, except the crossmatch to Gaia, where a radius of 0.5 arcseconds is used. A preliminary set of selections is made based on the quality of the data in PanSTARRS and 2MASS. The cuts on PS1 data require at least 2 good detections out of the 5 PS1 bands, on the basis of the \texttt{nmag\char`_ok} parameter and that the difference between the reported PSF magnitudes and Aperture magnitudes be less than 0.1 in at least 2 bands. The procedure further requires `good detections' in at least 4 out of 8 bands across both surveys and at least 2 good detections in PS1. The exact quality criteria that define `good detections' for each survey are outlined in Appendix A.1.

The cuts most relevant to the question of separating extragalactic objects from true stars include the PSF-Aperture cut for PS1, and the \texttt{gal\char`_contam} and \texttt{ext\char`_key} based criteria for defining `good detections' for 2MASS. The first of these selections is based on the observation that PSF photometry tends to underestimate the flux from extended objects like galaxies relative to the aperture photometry of extended objects \citep[e.g.,][]{strauss2002spectroscopic}. The relevant 2MASS based criteria flag extended objects or those that lie within the 20 mag arcsecond$^{-2}$ elliptical isophote of an extended source. The specifics of these cuts are outlined in Appendix A.1.

This catalog of objects consists of magnitudes in eight bands (five for PS1 and three for 2MASS), as well as parallax information from Gaia EDR3. This information is used in deriving per star posteriors on distance modulus and reddening for each source.

\textit{Stellar Posterior Inference: } We use the same stellar posterior inference framework as was used for the \bay{19} dustmap, but with parallax measurements from Gaia EDR3. The stellar posterior inference framework only uses parallax information for `high fidelity' Gaia detections. We define high fidelity detections as those which have a Renormalized Unit Weight Error (RUWE) less than 1.4 and an `astrometric fidelity' prediction greater than 0.5. The fidelity prediction for each source is obtained by querying the table generated by the astrometric fidelity classifier \citep{vo:gedr3spur_main, rybizki2022classifier}, a neural network that predicts the fidelity of a source conditioned on input features such as the RUWE, \texttt{astrometric\_excess\_noise}, parallax and proper motion measurements, among others. 

The stellar posterior inference framework is described in \cite{green2014measuring}, \cite{Green2015}, and Section 2.1 and Appendix A in \cite{Green2019}, and summarized here. The observed magnitude of an object in any band $\hat{m}_b$ is modeled as the sum of the distance modulus, the intrinsic magnitude and the extinction in that band, with an additional noise term. The $\hat{}$ symbol is used to denote observable quantities, which are the magnitude in each band $\hat{m}_b$ and the parallax $\hat{\varpi}$.
\begin{align}
& \hat{m}_b \sim \mathcal{N} (m_{b,th}, \hat{\sigma}_{m(b)}^2) \nonumber \\
& \text{where, } m_{b, th} = M_b(\Theta) + \mu + A_b \\
& \text{and, } A_b = R_bE \nonumber \\
& \implies \hat{m}_b - M_b(\Theta) \sim \mathcal{N}(\mu + R_bE, \hat{\sigma}_{m(b)}^2) \label{eqn: photogen}
\end{align}
$R_b$ is the component of the reddening vector $\vec{R}$ that converts the wavelength-independent reddening to the extinction in band $b$. As in \cite{Green2019}, we choose a reddening vector typical of the diffuse interstellar dust.\footnote{
We use the prescription of \cite{schlafly2016optical}, who define a parameter $R'_V$ that is similar to the usual $R_V\equiv A_V/E_{B-V}$.  A value of $R'_V=3.3$ corresponds to $R_V=3.1$, typical of dust in the diffuse ISM.}  Since the magnitudes are linear in extinction and distance modulus, and the photometric errors approximately Gaussian, the maximum likelihood distribution of the distance modulus and extinction, given stellar type are Gaussian (Equation \ref{eqn: photogen}). The maximum likelihood distance moduli and extinctions are calculated for an ensemble of stellar types, and the likelihood of the data given the stellar type and the maximum likelihood values. This comprises a Gaussian Mixture Model where the components (i) correspond to each stellar type, with the means, weights and covariances given by:
\begin{align}
&   \text{Mean}_i = [\mu_{i}, E_{i}] \nonumber \\
&   w_i = p(\hat{m}, \hat{\varpi} | \Theta_i, \mu_{i}, E_{i}) p(\Theta_i, \mu_{i}, E_{i}) \\
&   C_i = C = (A^T \Omega A)^{-1} \nonumber \\
& \text{where, } A = \big[\vec{1},  \vec{R}\big]_{|\vec{m}|\times2} \text{ and } \Omega = \mathcal{D}(\sigma_m^{-2}) \nonumber
\end{align}
The $16^{th}, 50^{th}, 84^{th}$ percentiles of this distribution were computed and then converted to a median and standard deviation of the distance modulus and extinction.

\textit{Bayestar Stellar Models:} 
The model described above requires the absolute magnitude in different bands as a function of stellar type $M_b(\Theta)$. The stellar type, $\Theta$ is parameterized by the absolute magnitude in the PS1 \textit{r} band $M_r$, and the metallicity [Fe/H]. The prescription for deriving $M_b(\Theta)$ is described in detail in \cite{Green2015, Green2019} and involves fitting a stellar locus in 7-D color space to a set of around 1 million stars in low reddening ($E(B-V)_{SFD} <0.1$ mag) regions. This locus is then associated with an absolute magnitude and a metallicity using \cite{ivezic2008milky} to yield an ensemble of stellar models $\{\vec{M}(M_r, [Fe/H])\}_i$.

In contrast to \bay{19}, we aim to make a 2-D map with cosmological applications in mind.  A 2-D map contains far fewer degrees of freedom to constrain, so we can afford to be somewhat conservative in our selection of stars. We want a sample of high purity that lies behind most of the Milky Way dust.

\subsection{Secondary Selections}\label{subsec:cuts}
\begin{figure}
\centering
    \includegraphics[keepaspectratio=true, width=\linewidth,height=200pt]{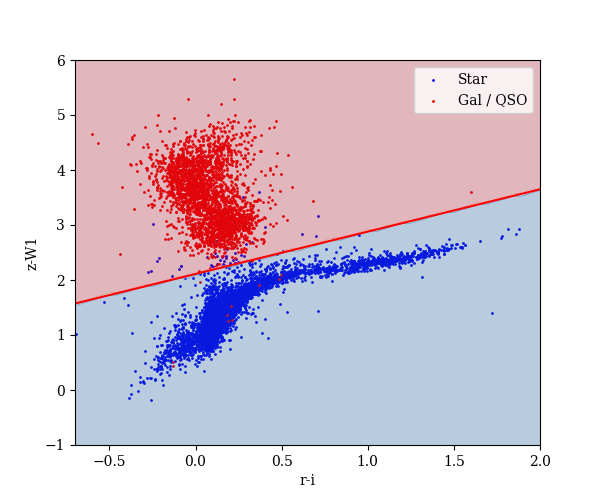}    \includegraphics[keepaspectratio=true, width=\linewidth,height=200pt]{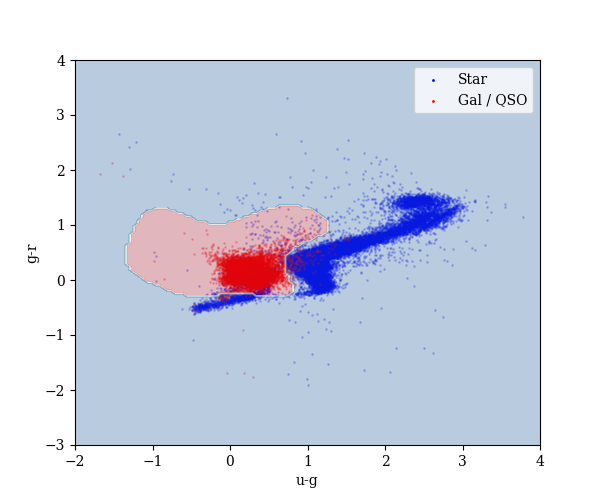} 
\caption{Decision boundaries for the PanSTARRS-WISE r-i, z-W1 cut (upper) and the SDSS u-g, g-r cut (lower).
\label{fig:models}}
\end{figure}
The combination of secondary stellar selections used in this paper is described below: \newline
We retain in our sample only objects that pass the following cuts:
    \begin{itemize}
        \item $d_{50}\sin(\textit{b})>400$pc, where $d_{50}$ corresponds to the distance (in pc) derived from the median of the object's posterior on distance modulus.
        \item $\sigma_{\mu}<1.5$ where  $\sigma_{\mu} = (\mu_{84} - \mu_{16})/2$ and $\mu_{P}$ is the $P^{\rm th}$ percentile of the object's posterior on distance modulus.
        \item the object passes a `linear' PanSTARRS-WISE color cut.
        \item the object passes a color cut using SDSS \textit{ugr}.
        \item the object has a valid parallax measurement in Gaia EDR3.
        \item the reduced chi-square $\chi^2_{r} < 3$.
    \end{itemize}
The first two cuts on distance modulus are based on our modelling assumptions. The third and fourth cuts, plotted in Figure \ref{fig:models}, are color cuts whose decision boundaries are trained on a subset of objects with spectroscopic labels from SDSS DR17 `SpecObj'. The third cut is a color cut using WISE and PanSTARRS \textit{riz} that retains objects that are identified as `stars' by a support vector machine trained on their $m_r - m_i$ and $m_z - m_{W1}$ values. This cut effectively classifies all objects with $[m_z - m_{W1}] < b + A\times[m_r - m_i]$ as stars, where $b = 2.11, A= 0.77$. Unlike most stars, which typically have their radiation peak specifically in the optical, QSOs and galaxies have extended emission spectra, and relatively higher flux in the infrared. A similar boundary was also used by \cite{Green2019} to assess the extent of quasar contamination, but not as a cut on objects. The fourth cut retains objects that are identified as `stars' by a support vector machine that uses the objects' $m_u - m_g$ and $m_g - m_r$ colors from the SDSS DR14 survey. This is motivated by the ultraviolet selection method: QSOs consist of Active Galactic Nuclei which often exhibit excess flux in the ultraviolet due to the blue bump. The training and implementation of these cuts is described in Appendix A2. The fifth cut, retains objects that possess a parallax detection in Gaia EDR3 that isn't a \texttt{NaN}. The sixth cut acts on the reduced chi-squared statistic of the object's posteriors and excludes sources that are a poor fit to the Bayestar stellar models. We sequentially examine how effectively each cut removes extragalactic objects in Section \ref{sec: specmatched}. While additional cuts on proper motion could theoretically increase the purity of objects included in the spectroscopically matched sample, we found that their addition did not seem to further reduce the angular cross-correlation of the resulting map, indicating that that information was possibly redundant with the other cuts described here.

\section{Reconstruction Schemes} \label{sec: reconstruction}
We choose to construct our maps at HEALPix \citep{gorski2005healpix} \ns{2048}, and with a point spread function FWHM $6.1'$ and $15'$. For all stars that pass the selections in Section \ref{sec:selections}, stars within 5 $l_{PSF}$ (where $l_{PSF} = FWHM/2.355$) of a given pixel are filtered, and weighted by the following expression: 
\begin{align}
& w_{\rm PSF}[p, s] = \exp \left[\frac{-(\vec{r}_p - \vec{r}_s)^2 }{2l_{\rm PSF}^2}\right]\\
& w_{\rm inv}[p, s] = \frac{w_{\rm PSF}}{\sigma_s^2 + \sigma_{\rm ref}^2} \\
& w_{\rm eff}[p, s] = \frac{w_{\rm inv}}{\sum_{s} w_{\rm inv}} \\
& {\rm Mean}(E[p]) = \sum_{s} w_{\rm eff}[p, s]*E_{50}[s] \\
& {\rm Var}(E[p]) = \sum_{s} w^2_{\rm eff}[p, s]*\sigma^2_{E}[s]  \label{eqn:recon_sigma}\\ 
& \text{where, }  \frac{1}{\sigma^2_{\rm ref} + \sigma^2_{\rm 1\%}} = \frac{10}{\sigma^2_{\rm ref} + \sigma^2_{\rm 99\%}}
\end{align}
$\sigma_{\rm ref}$ is set such that the inverse variance weighting factor assigned to a star with a posterior extinction uncertainty corresponding to the $1^{st}$ percentile of stars which passed the cuts in Section \ref{sec:selections} in the parent \ns{32} pixel is 10 times larger than an equivalent star at the same distance but with a posterior extinction uncertainty corresponding to the $99^{th}$ percentile. Since the maps' reddenings are calibrated to SFD, we multiply the maps by 0.856, to convert to $E(B-V)$. As a postprocessing step, a constant is added to each map such that the resulting maps' means match the mean of the extinction values predicted by SFD (also multiplied by 0.86) over the patch of the sky at $b>60^\circ$. This postprocessing step does not affect the angular cross-correlation plots in Figure \ref{fig:acc_main} since the mean of the maps is subtracted as a preprocessing step before computing the estimate. The 0.86 factor multiplying SFD was derived in \cite{schlafly2010blue} by recalibrating SFD with respect to the blue tip of the stellar locus using SDSS photometry. All comparisons with SFD in this work use this factor.

\section{Analysis}\label{sec:analysis}
We use two figures of merit to assess the extent of extragalactic contamination in our maps.

The first test of how well our selection eliminates galaxies and quasars from the catalog sample is to examine a subset of objects which have spectroscopic labels, and assess the extent of contamination from extragalactic objects before and after applying the selection. The subset of objects with spectra is only a small fraction ($\sim 1\%$) of the full set of objects whose reddening posteriors are used in the reconstruction. This testbed allows us to directly examine the effect of adding or removing different cuts. The spectroscopically matched sample is not intended to be representative of the distribution of objects in our the full catalog. The spectroscopically matched sample is likely to contain more QSOs, since the distribution is reflective of the objects that were targeted to have their SDSS spectra measured.

The second, complementary test examines the spatial correlation of a dust map with LSS as traced by extragalactic objects with redshifts from the Sloan Digital Sky Survey (SDSS).  For this we use the clustering-redshift based method \citep{chiang2019extragalactic}, originally introduced in \cite{menard2013clustering}. This technique has been applied to infer the correlation with redshift of an unknown discrete sample or a continuous field with reference objects with known redshifts. This can be used to infer the redshift distribution of a sample of objects, or, as in \cite{chiang2019extragalactic}, evaluate whether extinction patterns in maps correlate with the positions of objects in specific redshift bins. Our implementation of the measurement differs from their approach in certain respects, specifically in terms of how we measure the error on the cross-correlation signal, the smoothing scale used to preprocess the maps, the level of pixelization at which input maps are queried and the absence of an inverse variance weighting scheme. 

We also examine the extent of deviations from SFD and the level of noise as a function of latitude.

\subsection{Performance on a spectroscopically matched sample} \label{sec: specmatched}
\begin{figure}[!h]
\centering
    \includegraphics[keepaspectratio=true, width=\linewidth,height=300pt]{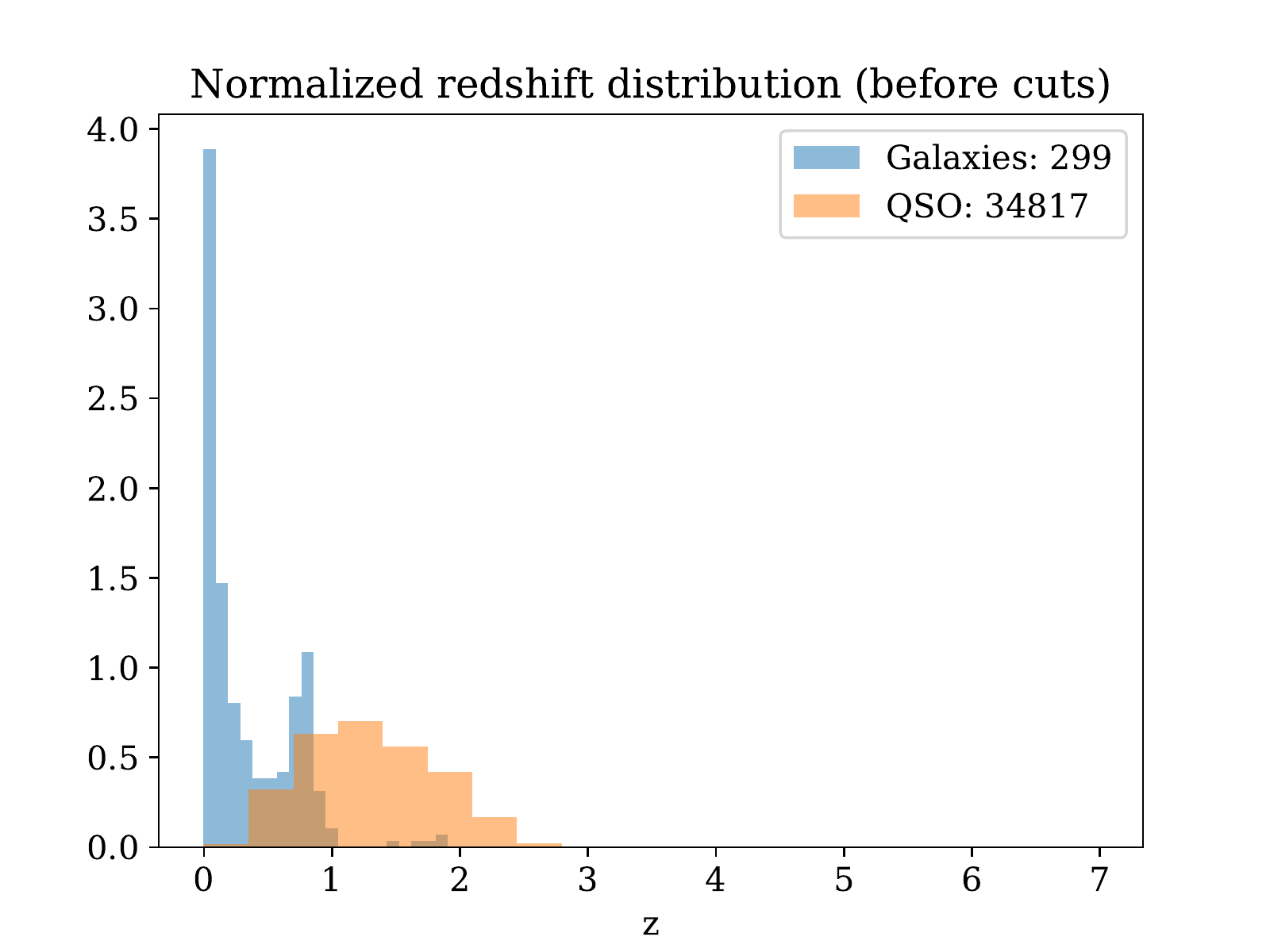}    \includegraphics[keepaspectratio=true, width=\linewidth,height=300pt]{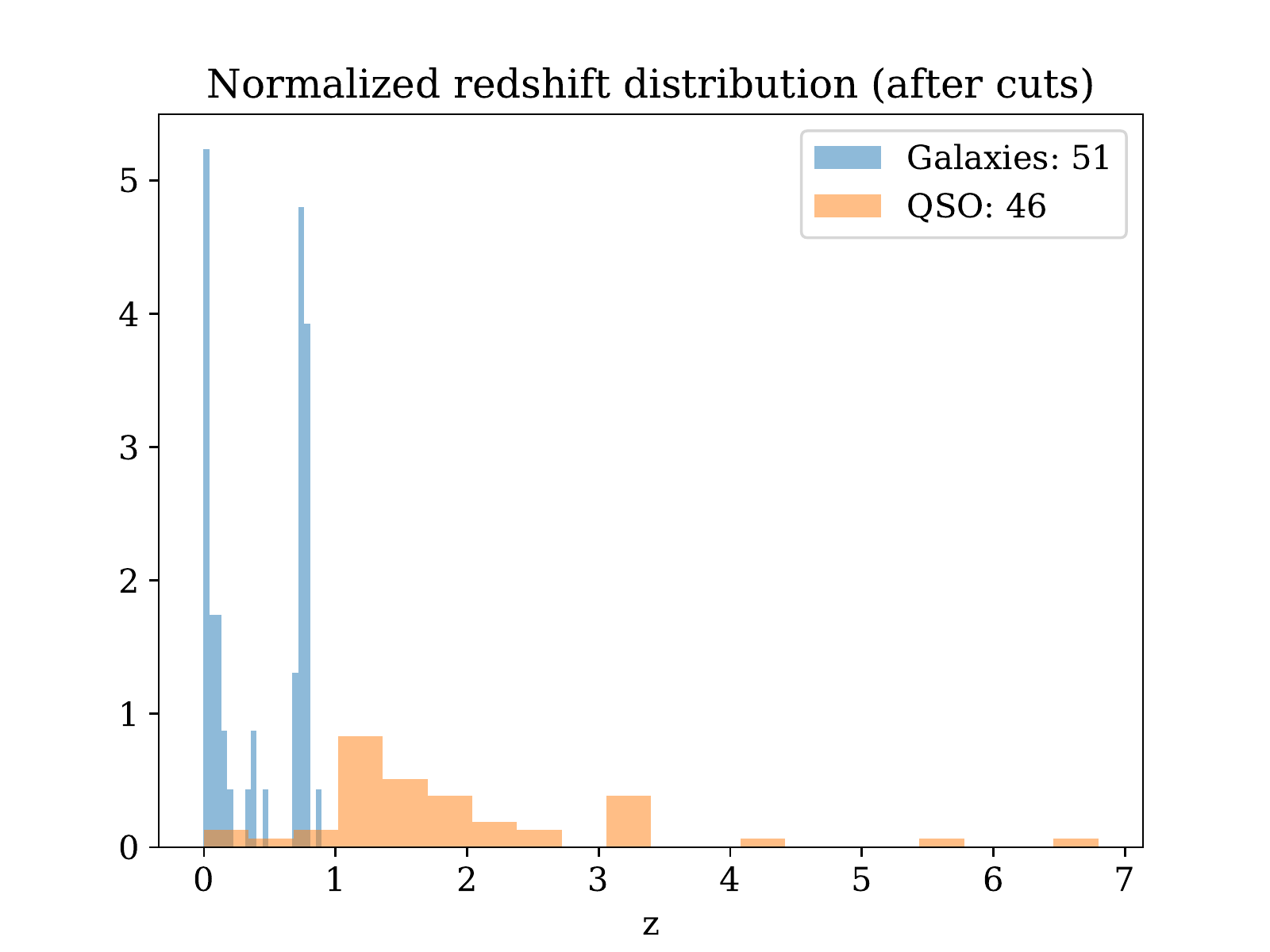}
\caption{Redshift distribution of extragalactic objects in the spectroscopically matched subset in the Northern Galactic Hemisphere before and after secondary selections are applied.  Galaxies were mostly removed by the preliminary cuts.
\label{fig:zdbn}}
\end{figure}

To examine the efficacy of our secondary selections, we cross match stars from the catalog obtained after the preliminary set of selections (Section \ref{sec:prelim-selections}) for $b>40^{\circ}$ to the SDSS Data Release 17 `SpecObj' catalog, with a crossmatch radius of 1". To ensure the spectroscopic labels are reliable we further impose the following quality cuts to select objects with `reliable' spectra, and require \texttt{rchi2}$<2$, \texttt{chi68p}$<2$, and \texttt{sn\char`_median\char`_all}$>10$. This yields a set of 210713 objects, of which 175597 (83\%) are stars, 34817 (17\%) are QSOs, and 299 are galaxies (0.14\%). The low number of galaxies in the sample without any secondary selections applied can be attributed to the aperture minus PSF magnitude cut applied in the preliminary selections. The color cuts involving SVM models were trained on a subset of the spectroscopically matched sample from the Northern Galactic Hemisphere. After applying all cuts, 138797 stars, 46 QSOs and 51 galaxies remained in the sample. Figure \ref{fig:zdbn} plots the normalized redshift distribution of objects in the spectroscopically matched sample before and after the cuts are applied. While high $z>3$ objects account for a small fraction of extragalactic objects before making cuts, they account for a \textit{relatively} larger fraction of objects that remain in the sample after applying the stellar selections. Tables \ref{tab:spec_classes-test} and \ref{tab:ablation-test} examine the behavior of the cuts on an independent `test' set of spectroscopically matched objects with reliable spectra in the Southern Galactic Hemisphere, with  $b<-40^{\circ}$. None of the objects in this set was used to train the color cuts. The equivalent tables for the Northern Galactic Hemisphere are in the Appendix (Tables \ref{tab:spec_classes}, \ref{tab:ablation}). Before applying any secondary selections, this set contained 55474 (85\%) stars, 9816 (15\%) QSOs, and 87 (0.13\%) galaxies. After applying the stellar selections, 45473 stars (99.89\%), 30 galaxies (0.06\%) and 16 QSOs (0.04\%) are left. We examine the incremental effect of each cut one at a time on the sample, and perform ablation tests where the effect of excluding each cut are tested. The cuts on distance modulus have negligible effect on the purity of the sample, while the SDSS \textit{ugr} and PanSTARRS-W1 cuts are particularly valuable in terms of removing QSOs. The Gaia EDR3 parallax detection cut had the weakest effect overall but was included because it appeared to reduce contamination from galaxies better than some of the other cuts as well as reduce noise in the resulting map.

\begin{deluxetable*}{cccc}
\tablenum{1}
\tablecaption{Distribution of objects from the spectroscopically matched sample in the Southern Galactic Hemisphere in different classes as stellar selection cuts are incrementally applied. The number reports the number of objects in each class. The figure in brackets reports the fraction of objects belonging to a certain category after a particular choice of cuts is applied $\frac{n^{After Cuts}_{Class}}{n^{After Cuts}_{Total}}$. Each row in the table also has all the cuts in the rows preceding it, applied. (The percentages in each row sum to 100\%). \label{tab:spec_classes-test}}
\tablewidth{0pt}
\tablehead{
\colhead{Cuts} & \colhead{Stars: No./ Proportion (\%)} & \colhead{QSOs: No./ Proportion (\%)} & \colhead{Galaxies: No./ Proportion (\%)} 
}
\startdata
No Cuts Applied & 55474 (84.85) & 9816 (15.01) & 87 (0.13) \\
+$d_{50}\sin(b) > 400{\rm pc}, \sigma_{\mu}<1.5$ & 50205 (84.69) & 8992 (15.17) & 84 (0.14) \\
+WISE nondetection/($r-i, z-W1$) cut & 49464 (98.95) & 468 (0.94) & 58 (0.12) \\
+ eDR3 parallax detection & 49305 (98.96) & 468 (0.94) & 48 (0.10) \\
+SDSS $(u-g, g-r)$ cut & 48998 (99.82) & 48 (0.10) & 41 (0.08) \\
+$\chi^2_r<3$ & 45473 (99.90) & 16 (0.04) & 30 (0.06) \\
\enddata
\end{deluxetable*}

\begin{deluxetable*}{cccc}
\tablenum{2}
\tablecaption{Ablation Tests on the spectroscopically matched sample in the Southern Galactic Hemisphere: All cuts in the Secondary Stellar Selection EXCEPT a given cut are applied.  \label{tab:ablation-test}}
\tablewidth{0pt}
\tablehead{
\colhead{Excluded Cut} & \colhead{Stars: Number} & \colhead{QSOs: Number} & \colhead{Galaxies: Number } 
}
\startdata
No cuts excluded & 45473 & 16 & 30 \\
\hline
-WISE nondetection/(r-i, z-W1) cut & 45731 & 37 & 31 \\
-eDR3 parallax detection & 45616 & 16 & 34 \\
-SDSS (u-g, g-r) cut & 45612 & 71 & 32 \\
-$\chi^2_r<3$ & 48998 & 48 & 41 \\
\enddata
\end{deluxetable*}

\subsection{Measuring Correlations with Large Scale Structure} \label{sec: acc}
We examine the angular cross correlation of extragalactic reference objects in different redshift bins with two existing emission-based maps, two existing extinction-based maps,  and the new maps we construct, using the clustering-based redshift estimation technique.

We briefly summarize dust mapping efforts below, and the maps we compare against. The Burstein-Heiles map \citep{burstein1978hi} was one of the earliest attempts at dust mapping, and fit extinction as a function of galaxy counts and the column density of neutral hydrogen (H\textsc{i}). The Schlegel-Finkbeiner-Davis (SFD) map \citep{Schlegel1997}, generated a map of dust emission at 100$\mu$m using measurements from IRAS and DIRBE \citep{hauser1998cobe} with an angular resolution of FWHM = $6.1'$, after subtracting zodiacal light emission from the DIRBE maps and removing striping artifacts and point sources from IRAS. This emission map was converted to optical depth at 100~$\mu$m using a color temperature derived from the ratio of 100$\mu$m to 240$\mu$m emission in the DIRBE maps. The dust column density map was converted to optical reddening map by calibrating against the reddening of a sample of elliptical galaxies. The maps used by SFD have an estimate of the mean cosmic IR background subtracted, but not its anisotropy, leaving a residual correlation with large-scale structure. The extinction values from SFD are multiplied by the 0.86 factor derived by \cite{schlafly2010blue}.

The Planck 2016 products (\cite{aghanim2016planck}) include dust emission maps at 353, 545 and 857 GHz. The inputs to the map-generation procedure were the Planck temperature full-mission sky maps from the Planck 2015 data release (PR2) (\cite{planck2016planck}, \cite{adam2016planck}) in nine frequencies ranging from 30 to 857 GHz, as well as a combined temperature map at 100 $\mu$m based on a combination of the IRIS map (\cite{miville2005iris}) and SFD. The map has a spatially varying effective beam size depending on the signal to noise ratio over the sky and has a FWHM of 5' over 65\% of the sky but up to 15'-21' at the highest galactic latitudes. A dust extinction map is obtained by multiplying the dust optical density map at 353 GHz by a factor derived from the correlation between the reddening of quasars and dust optical depth (\cite{abergel2014planck}). 

 In recent years, several stellar-reddening based maps have been produced, differing in terms of their statistical reconstruction methods, volume covered, extents of discretization and choices of stellar posteriors eg: \citep{Green2015, kh2017inferring, Green2018, Green2019, lallement2019gaia, Leike2020}. The \bay{17} map \citep{Green2018} used photometry from Pan-STARRS 1 \citep{PS1_paper1} and 2MASS \citep{skrutskie2006two} to generate a stellar-reddening map in three dimensions for $\delta> -30^{\circ}$. This work used the reddening vector derived by \cite{schlafly2016optical} from the measurements of the dust extinction curves of thousands of stars with spectra from the APOGEE spectroscopic survey \citep{majewski2017apache} and ten photometric bands in the optical and infrared. The \bay{19} \citep{Green2019} map broadly adopted the same approach to deriving a three dimensional dust distribution, but differed in some key aspects, including the addition of distance constraints in the form of parallaxes from the Gaia Data Release 2 \citep{brown2018gaia}, a Gaussian Process based prior to regularize the spatial dust distribution and the discretization of the incremental extinction to multiples of 0.01 mag. We use the \textsc{dustmaps} package to query various dust maps \citep{2018JOSS....3..695M}. The reddenings from the \bay{} maps were multiplied by a factor of 0.856 to convert to units of $E_{B-V}$. We examine the extent of cross-correlation of two emission-based maps: SFD and the Planck 2016 map, hereafter denoted by GNILC, two stellar-reddening based maps integrated to the last distance bin (\bay{17} and \bay{19}) and the maps we constructed at a FWHM of 6.1' and 15'.

\subsubsection{Implementation}
\newcommand\dele{\delta E}
Our implementation of the angular cross-correlation estimator is described below, and in Appendix \ref{sec:acc_deriv}:
\begin{enumerate}\setlength{\itemsep}{0pt}
\item Each input map $E(\phi)$ is queried at HEALPix \ns{2048}.
\item The input map is smoothed at $\sigma_{GAL}=30'$ to identify correlations arising from specifically extragalactic signals.  The dust density is, to first order, largely dependent on galactic latitude and to second order on filaments and structures in galactic dust. Thus, subtracting the input map smoothed at a scale $\sigma_{GAL}$ would largely eliminate these dependencies while preserving local fluctuations arising specifically from correlations with extragalactic structure. The mean of this difference is further subtracted to yield $\dele (\phi)$ (Equation \ref{eqn:delE}). 
\item A mask is defined that consists of the relevant area over which the cross-correlation signal is calculated. For the purpose of our analysis, our mask consists of the intersection of $b>50^{\circ}$ and the area where there are extragalactic tracers with which we can compute the cross correlation. 
\item We then evaluate the cross-correlation of the processed map with the fractional overdensity of reference objects at a given redshift, in each angular bin upto a certain angular scale set by $\theta_{max} =9.73'$ for each pixel (see Equation \ref{eqn:acc_z}). This gives us the angular cross-correlation signal in 34 redshift bins from $z=0.029-4.048$ for the input map.
\item To estimate the error on this signal, arising from the intrinsic noise of the map, we rotate the map about the north Galactic pole to 100 angles in the range $30^{\circ}-300^{\circ}$, and measure the root mean square deviation of the signal for the ensemble of rotated maps. In subsequent paragraphs, an $x$ `sigma' contour refers to $x$ times the RMSE computed in this fashion. The intuition underlying this step is that there is no physical reason a point on the sky would possess small-scale correlations with another random point on the sky -- such as that obtained by rotating the frame by an angle larger than some small angle.  Thus, in theory the signal for the rotated maps should be 0 at all redshifts. Any deviations from 0 are thus a measure of the contribution of random correlations between fluctuations in the map and the distribution of the reference objects to the estimate of the extragalactic correlation signal. Bootstrapping over the reference objects in the sample for each redshift bin underestimates the error bars relative to this approach to estimating the error on the correlation signal, as demonstrated in Section \ref{sec: ErrorBars}.
\end{enumerate}

\subsubsection{Error Bars}\label{sec: ErrorBars}
\begin{figure*}
\centering
     \includegraphics[keepaspectratio=true, width=.49\textwidth,height=800pt]{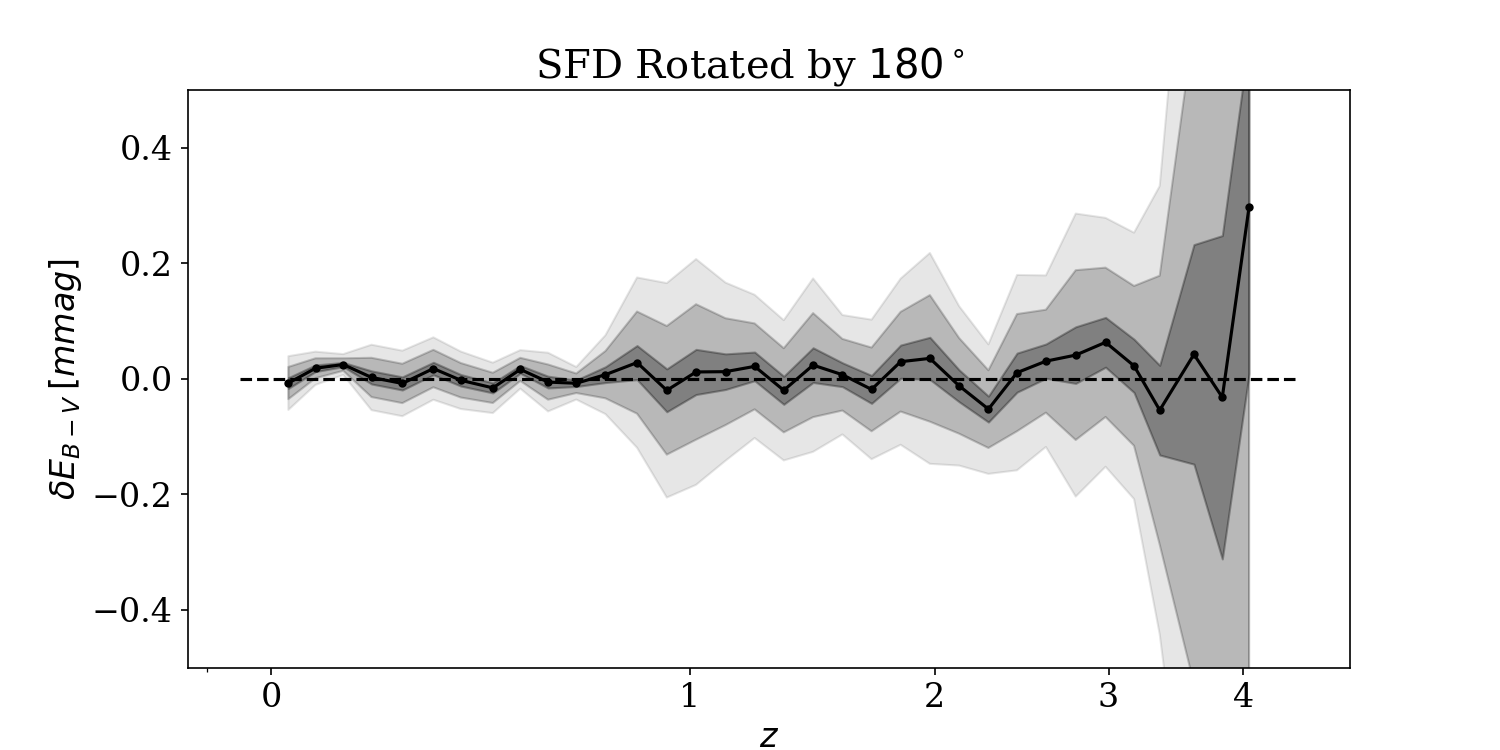}
    \includegraphics[keepaspectratio=true, width=.49\textwidth,height=800pt]{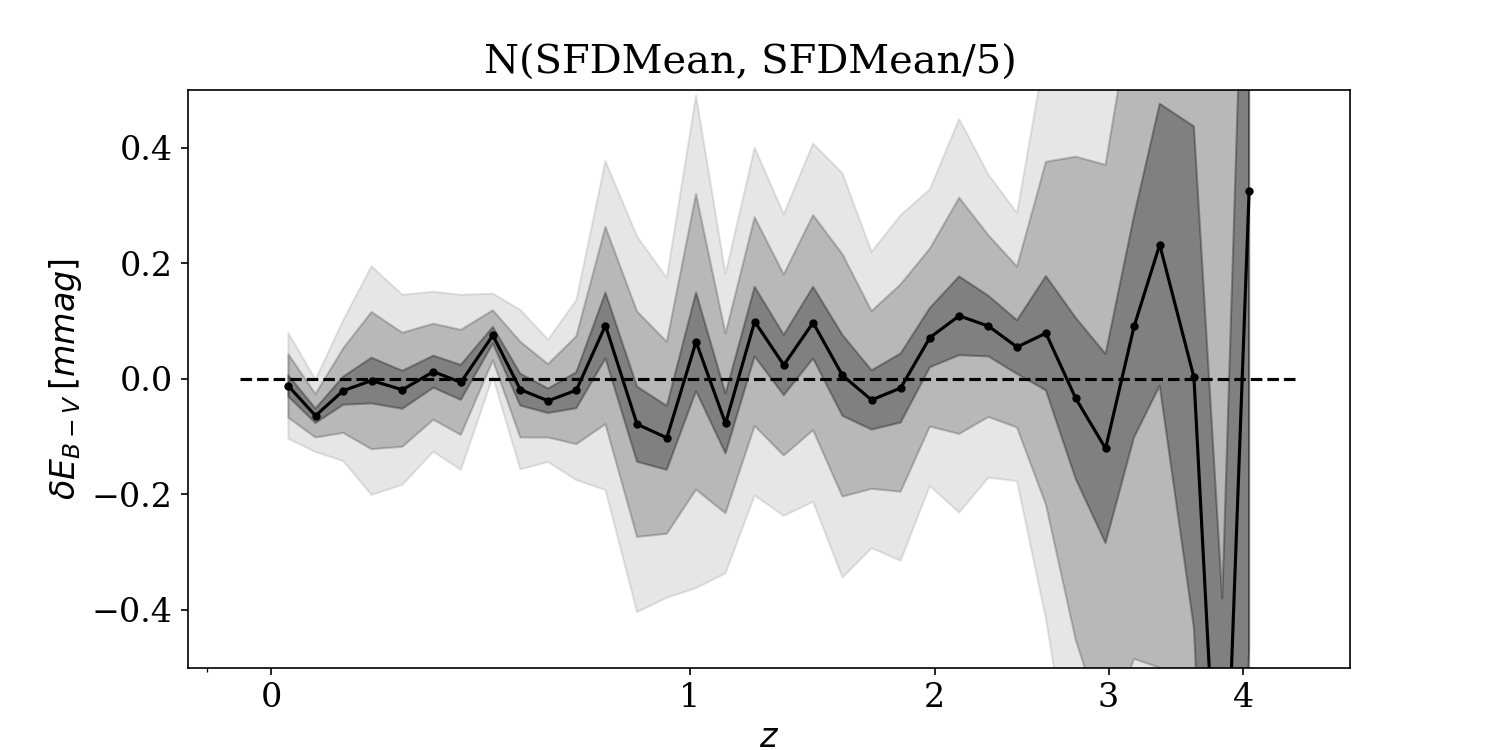} 
    \includegraphics[keepaspectratio=true, width=.49\textwidth,height=300pt]{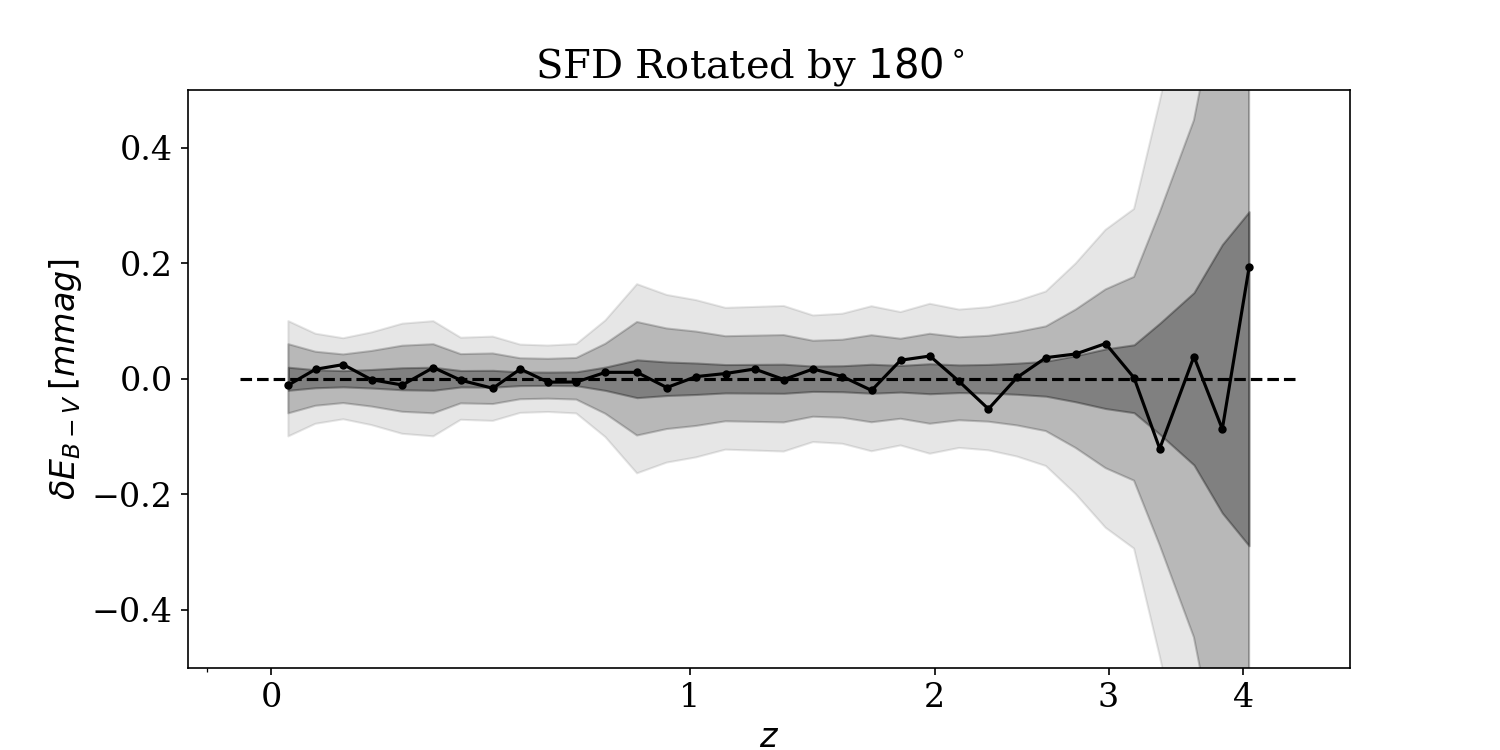}
    \includegraphics[keepaspectratio=true, width=.49\textwidth,height=300pt]{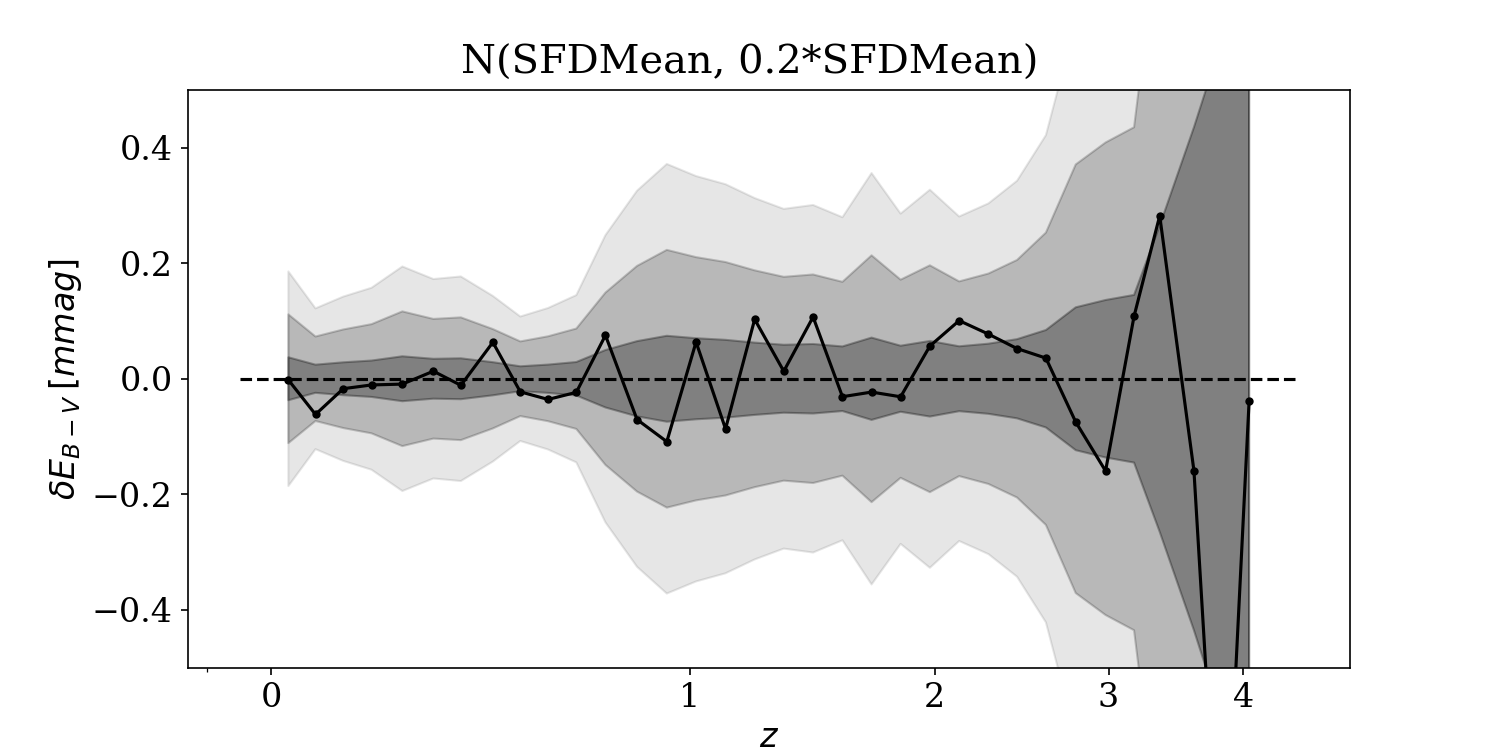}
\caption{Excess extinction in maps at the positions of reference extragalactic objects in redshift bins spanning z=0-4 for two null examples where no correlation should exist: a map of SFD rotated by $180^\circ$ (left column), and Gaussian noise (right column). The top row computes errors using the bootstrapped error bars, shown as grayscale regions around the measured correlation. The lower row computes errors using the rotation-based error bars, shown as grayscale regions centered on zero. The three envelopes correspond to 1, 3, and 5$\sigma$. 
\label{fig:null}}
\end{figure*}

Figure \ref{fig:null} depicts the angular cross-correlation signal for two null examples -- maps which we know should not possess any physical correlations with extragalactic structure. The null examples with the bootstrapped error bars, in the top panel seem to have more points that are around 5 sigmas off, such as the signal in the third redshift bin for SFD rotated by $180^\circ$. The level of signal measured using the rotation-based error bars does seem to lie largely within the 1 sigma contours. The differences in the choice of centering the error bars arises from the fact that the rotation-based errors account for deviations of the signal from 0 while the bootstrapped errors, account for deviations of the signal from its measured value, since we are bootstrapping over possible reference samples. For the majority of our analysis, we use the rotation-based error bars. Figure \ref{fig:acc_main_bootstrapped} in the Appendix also plots the angular cross-correlations for all the dust maps we consider using the bootstrapping-based errors.

\subsubsection{Comparisons}

\begin{figure*}[!h]
\centering
    \includegraphics[keepaspectratio=true, width=.49\linewidth,height=800pt]{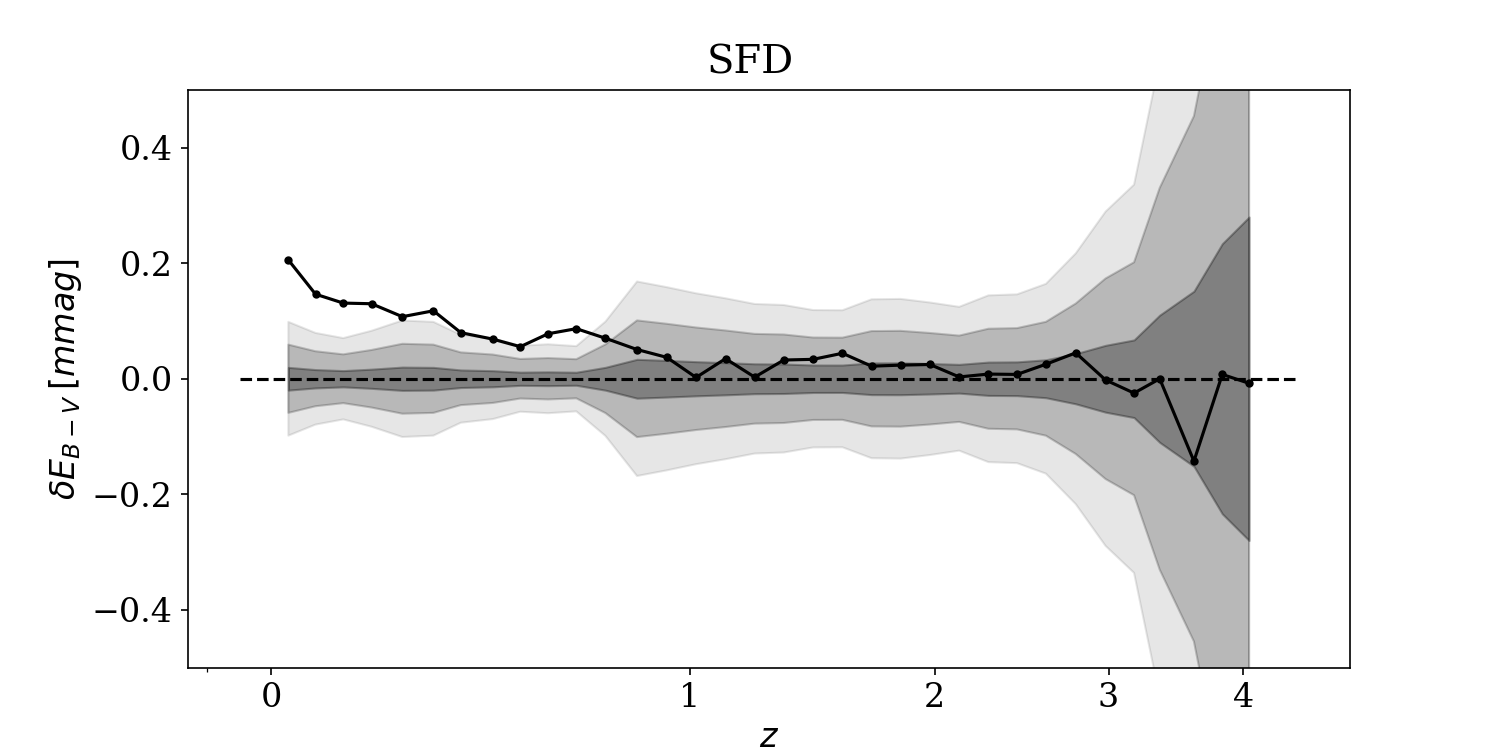} 
    \includegraphics[keepaspectratio=true, width=.49\linewidth,height=800pt]{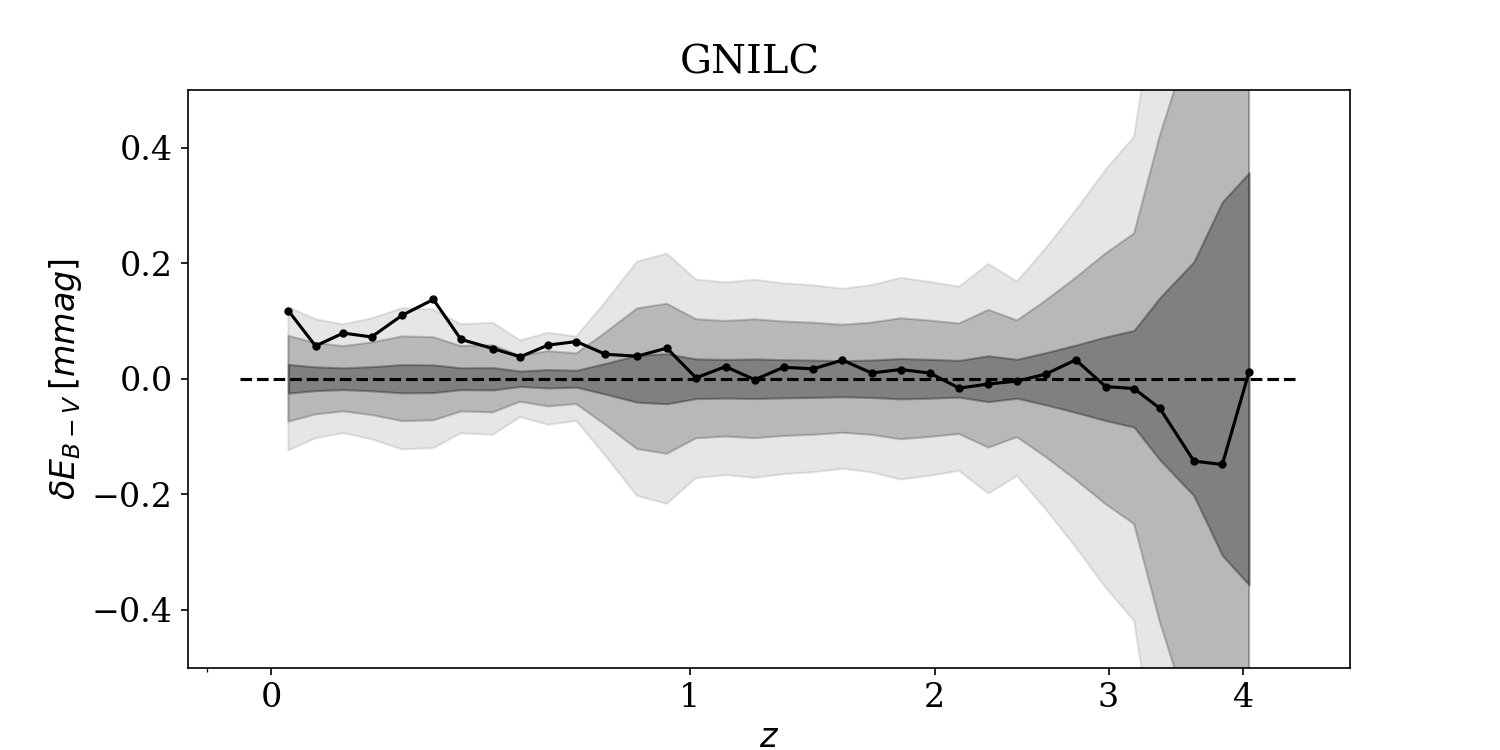}   \includegraphics[keepaspectratio=true, width=.49\linewidth,height=800pt]{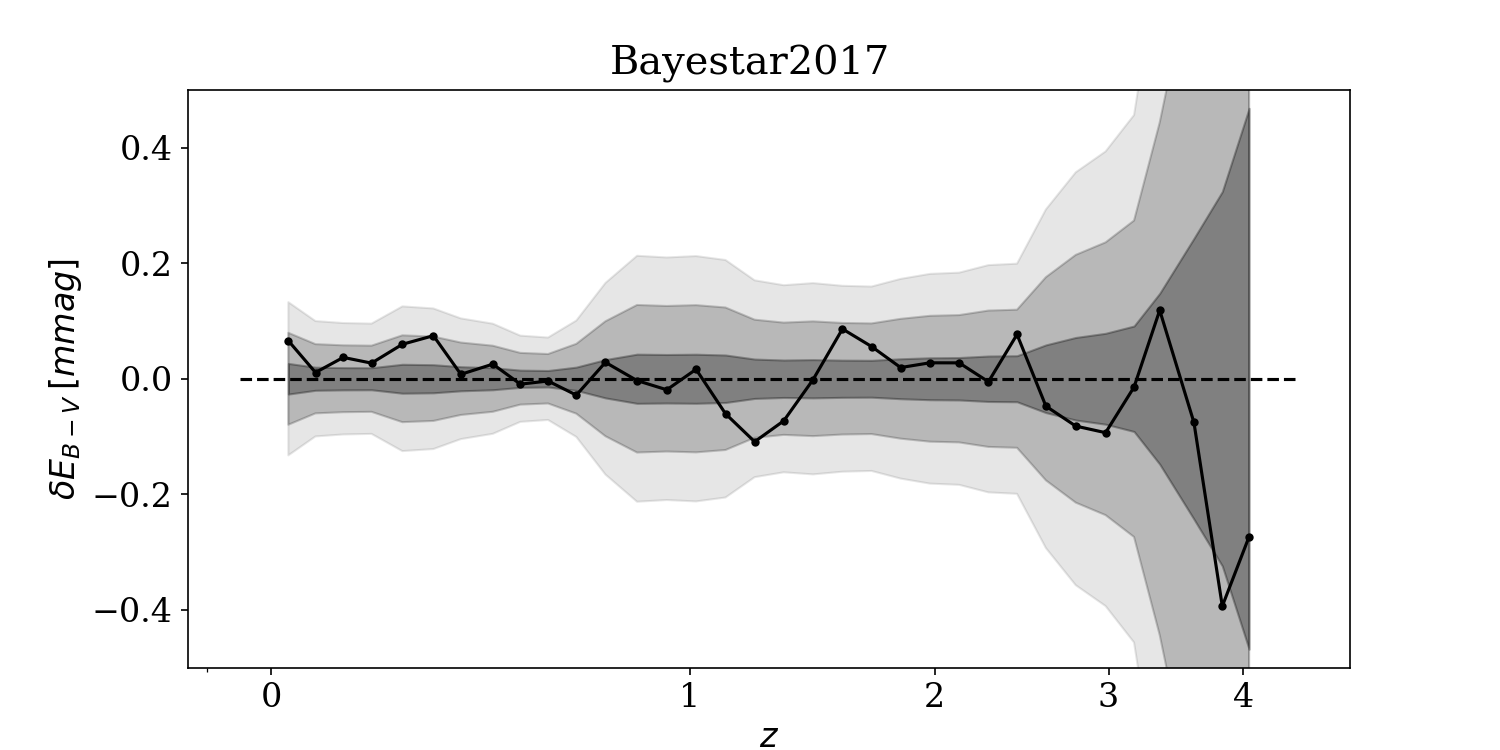}
    \includegraphics[keepaspectratio=true, width=.49\linewidth,height=800pt]{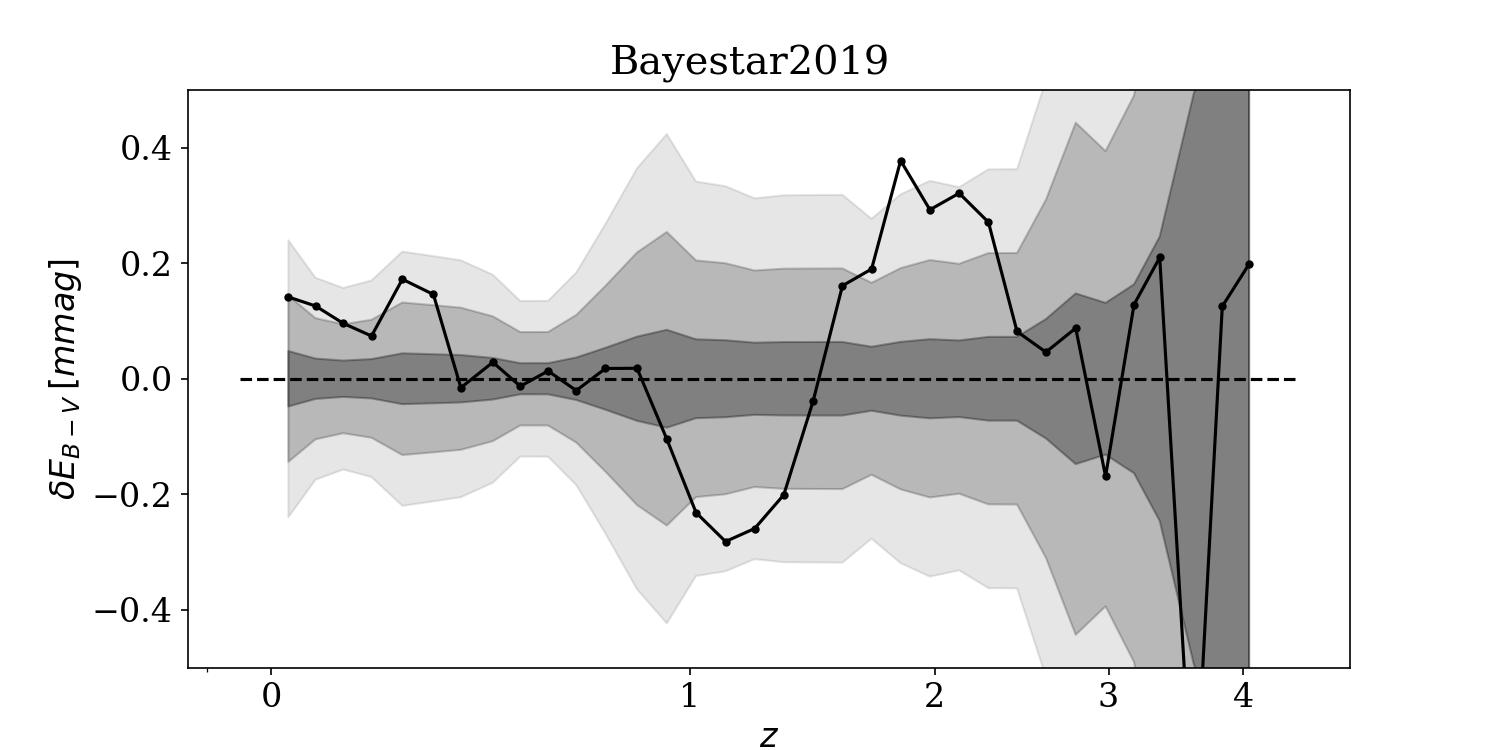}
     \includegraphics[keepaspectratio=true, width=.49\linewidth,height=800pt]{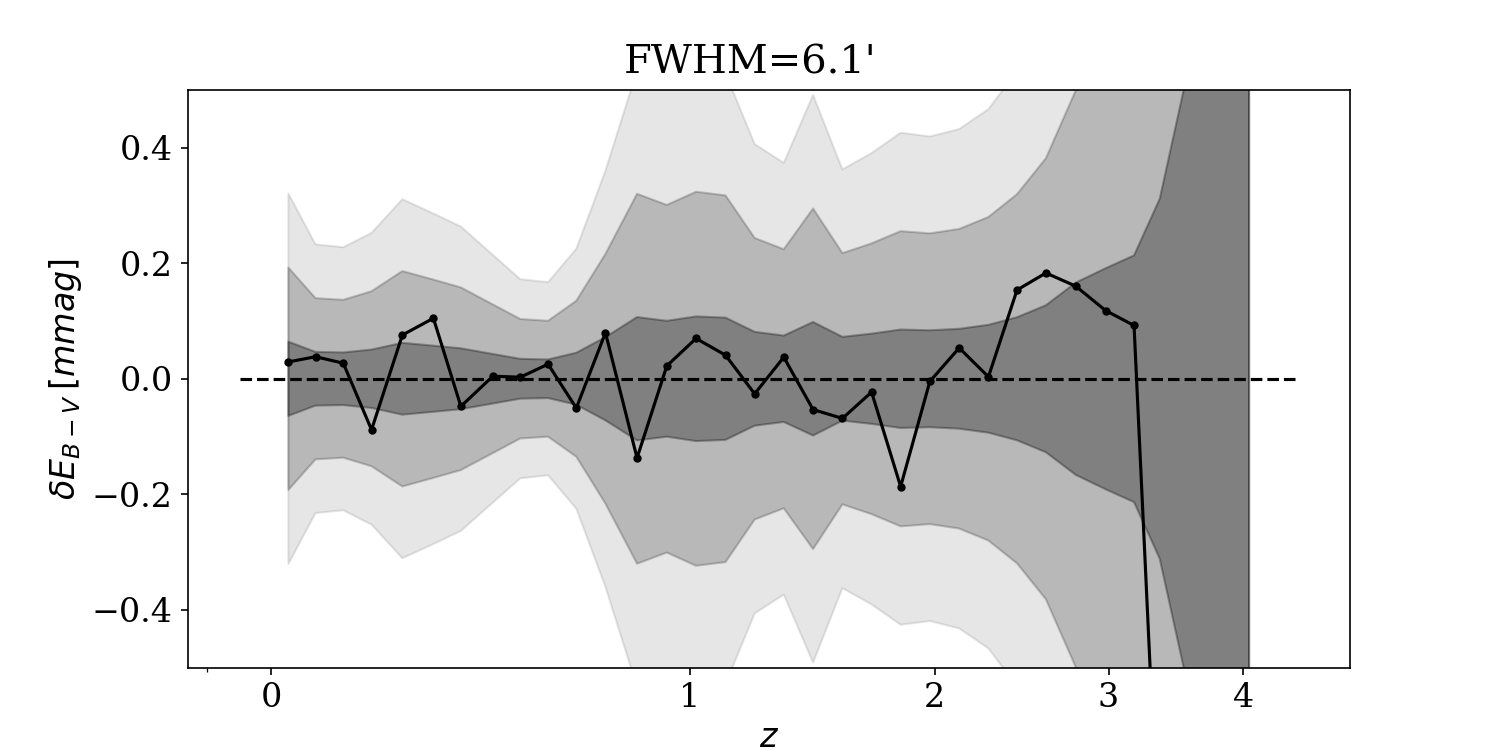}
    \includegraphics[keepaspectratio=true, width=.49\linewidth,height=800pt]{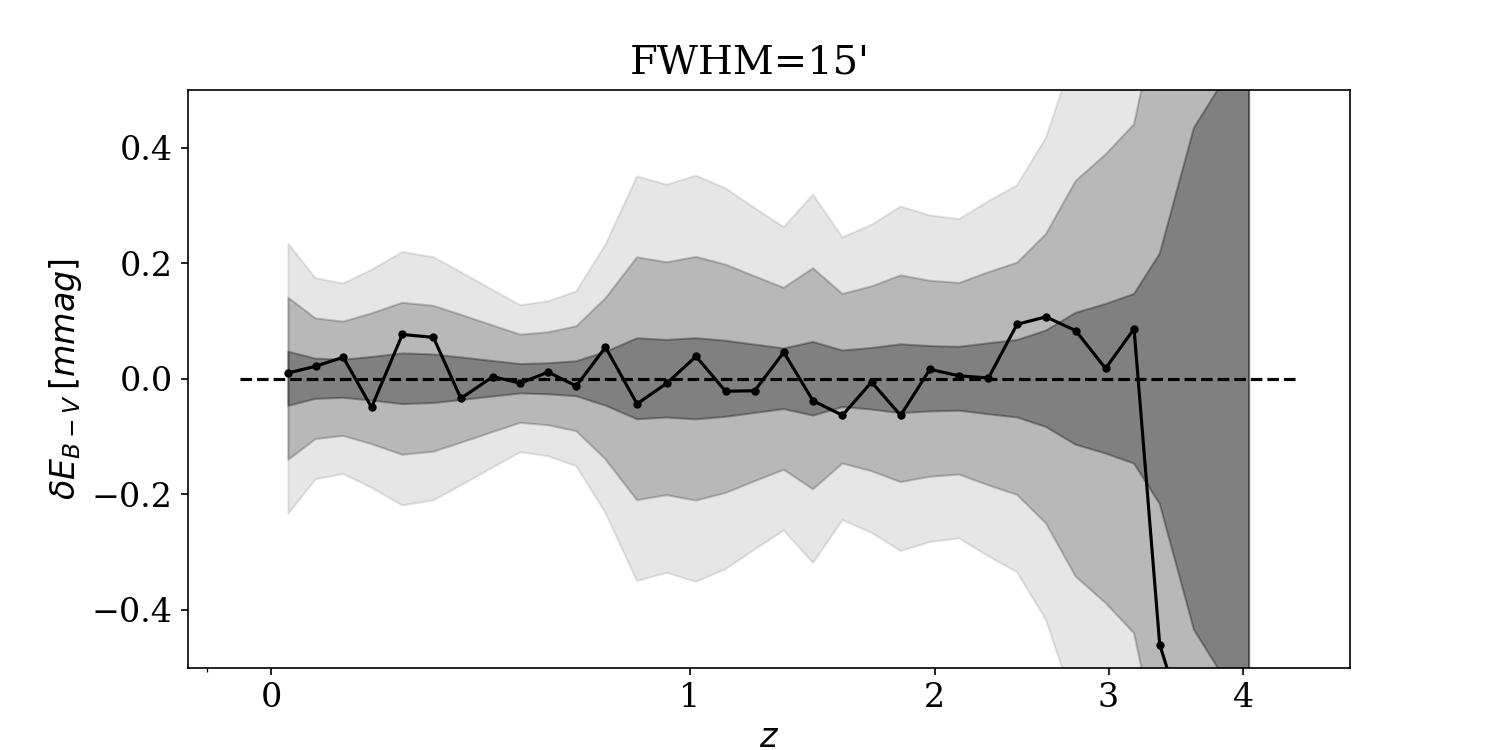}
    
\caption{Excess extinction in maps at the positions of reference extragalactic objects in redshift bins spanning $z=0-4$ for existing emission-based maps (top panel: SFD and the Planck 2016 GNILC dust emission map), existing stellar-reddening based maps (middle panel: \bay{17} and \bay{19}) and our reconstructions with FWHMs of 6.1’ and 15’. The sigma contours are obtained by rotating the map by 100 angles in the range $30^{\circ} - 300^{\circ}$ and calculating the root mean square error of the signals of the ensemble of rotated maps. For all maps, the signal was evaluated over the region $b>50°$.
\label{fig:acc_main}}
\end{figure*}

\begin{figure*}[!h]
\centering
    \includegraphics[keepaspectratio=true, width=.49\textwidth,height=300pt]{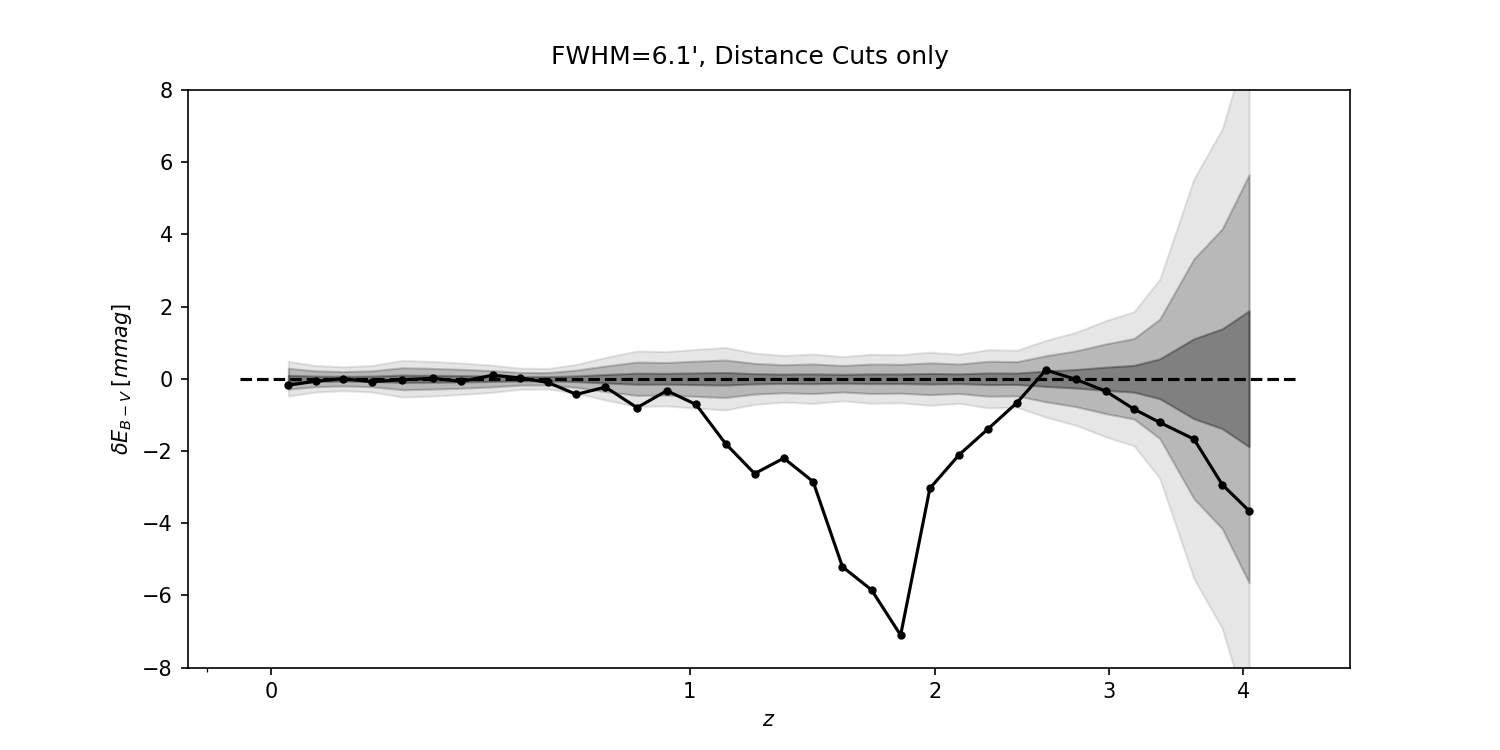}
    \includegraphics[keepaspectratio=true, width=.49\textwidth,height=300pt]{plots/accpanel/11-17_recon_6-1_final.png}
\caption{Angular cross-correlation signal in the 6.1' map. With only the distance cuts applied, the contamination from QSOs exceeds $5\sigma$ for $1<z<2.5$ (left). With the full set of stellar selections applied, the cross-correlation is broadly consistent with zero (right).    
\label{fig:ablationacc}}
\end{figure*}

We use this estimator, and the rotation-based error bars, to assess LSS contamination of two emission based maps -- SFD and GNILC, two existing stellar reddening based maps -- \bay{17} and \bay{19} and the maps we construct in Figure \ref{fig:acc_main}. Figure \ref{fig:acc_main_bootstrapped} depicts the same plots, along with the null example, with the bootstrapped error bars.
A positive correlation \textit{typically} indicates contamination from galaxies. In emission-based maps, this arises from imperfect removal of the cosmic infrared background, which leads to overestimated extinctions. Since it is foreseeable that the positions of galaxies in the reference samples are likely to correlate with sources of the cosmic infrared background (dusty star-forming galaxies), this leads to a higher cross-correlation signal at lower redshifts ($z<1$). We find that SFD's signal exceeds the 5 sigma contour at redshifts less than $z\sim1$. The GNILC map has a lower LSS correlation, with the signal within the 5 sigma level. As in \cite{chiang2019extragalactic}, the peak of the CIB-induced excess is affected by the K-correction and the wavelength that was used to measure dust emission.

In stellar reddening based dust maps, the redshift distribution of the angular cross correlation signal is less straightforward and would depend on what classes of extragalactic objects and at what redshift have the greatest impact on biasing extinctions. Of the existing stellar-reddening based maps, \bay{19}, had significantly higher correlations than both \bay{17} and the maps we construct. 
 The maps we construct (FWHM=$6.1'$ and FWHM=$15'$ in Figure \ref{fig:acc_main}) have correlations that are largely around the 1 sigma level, except in the highest redshift bins ($3<z<4$) where the signal is at the 3 sigma level and around $z\sim2$ in the FWHM=$6.1'$ map. Earlier stellar reddening based maps possessed negative correlations in the $z=1.1-1.3$ range, and positive correlations at the $z=1.6-2.2$ range at the $\sim 5$ sigma level in \bay{19} and the $\sim 1-3$ sigma level in \bay{17}. The correlated reddening deficit at around $z\sim1.2$ matches the increase in the QSO distribution observed in Figure \ref{fig:zdbn}. The noisier star-based maps have larger error contours since the noise in the map propagates into the signal's estimator, as in Figure \ref{fig:null}. It is also interesting that all maps appear to have underestimated signals at redshifts between $z=3-4$, however, the sparsity of reference objects at higher redshifts makes the contribution of the noise high too. The signal and the error bars both rise significantly at redshifts close to $z=3$ owing to the sparsity of the reference sample used to compute the signal at higher redshifts. In Figure \ref{fig:ablationacc}, we plot the angular cross-correlation of a map at FWHM=$6.1'$ without any of the extragalactic object-removal cuts applied and \textit{only} the distance based cuts. There is a significant negative correlation peaking near a redshift of $z=2$. After the cuts are applied, this $z=2$ correlation is now at less than 3 sigma.

\subsection{Comparisons to SFD}\label{sec:sfdcomp}
In this section, we examine how consistent our maps are with SFD, as a function of different extinction regimes and latitude.

\begin{figure*}
\centering
\includegraphics[keepaspectratio=true, width=0.3\linewidth,height=300pt]{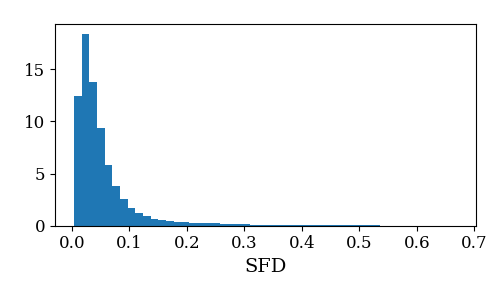}    \\
    \includegraphics[keepaspectratio=true, width=\linewidth,height=300pt]{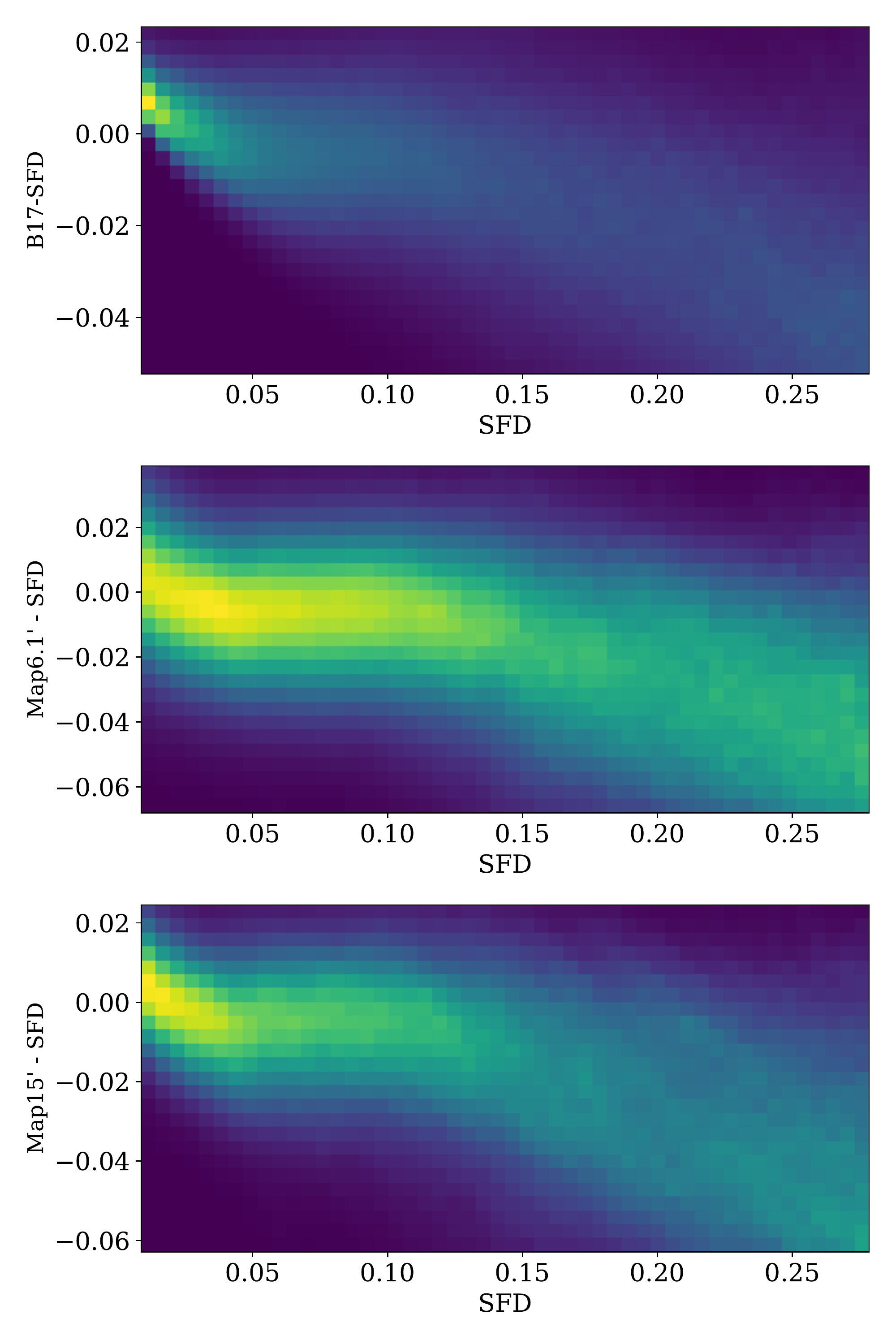}    
    \includegraphics[keepaspectratio=true, width=\linewidth,height=300pt]{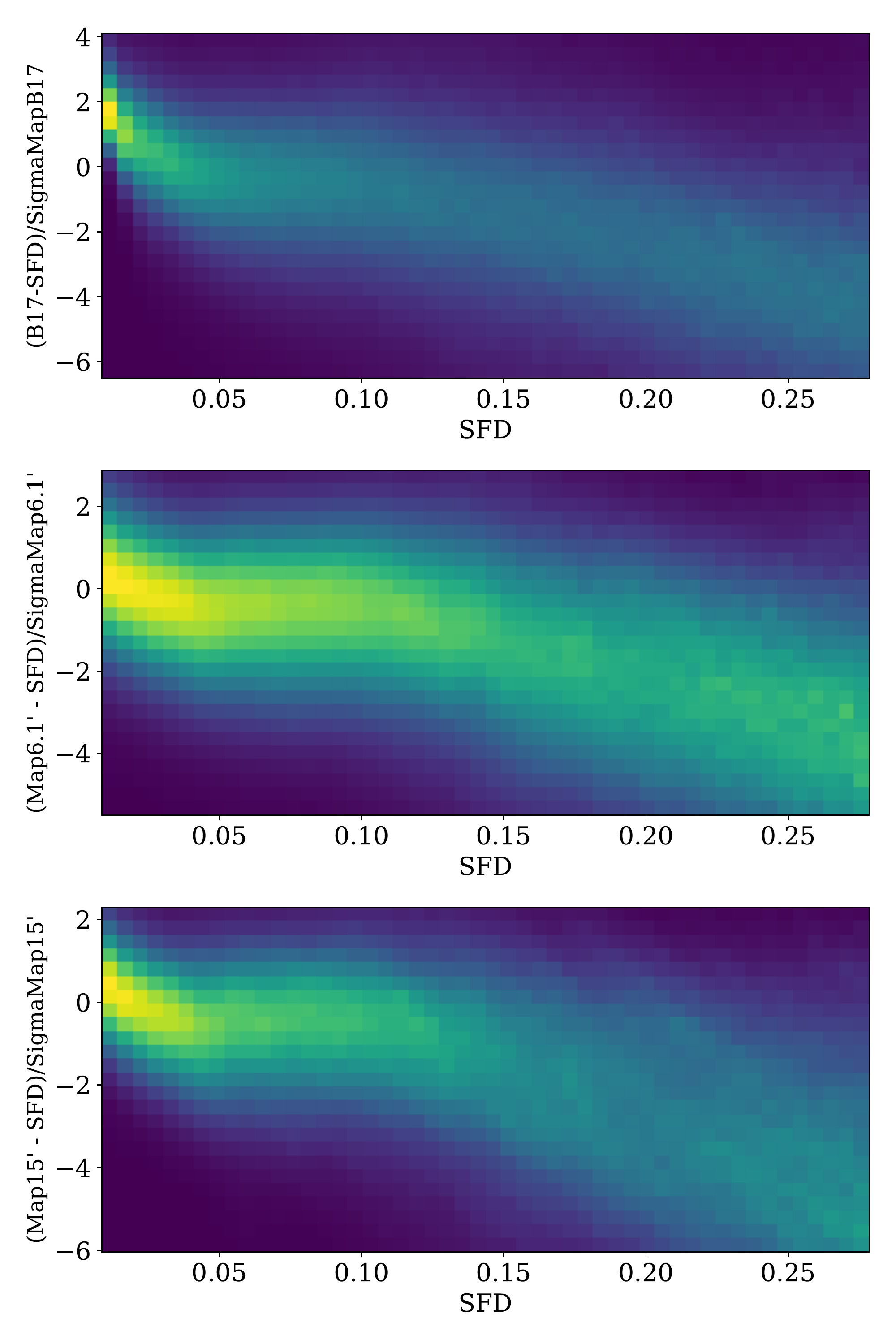} 
\caption{Histogram of SFD extinction values (top row). Two-dimensional histograms of the bias (left column) and uncertainty-normalized bias (right column) for \bay{17}, and our maps at FWHM=6.1' and FWHM=15' in the top, middle and bottom rows respectively. The zero point and uncertainty corrections have been applied here. The histograms are normalized such that the pixel values along the y axis for each value on the x axis (corresponding to a single SFD extinction bin) sum to 1.
\label{fig:hoggplot}}
\end{figure*}

\textit{Extinction}: We restrict ourselves to the footprint with valid predictions for our maps, and the \bay{17} and `19 maps -- the intersection of the Pan-STARRS1 footprint with regions with Galactic latitude $b>20^{\circ}$. The error propagation scheme we follow in Section \ref{sec: reconstruction} implicitly assumes uncorrelated errors, and as a consequence underestimates the error of the resulting map, particularly at lower latitudes and for the FWHM=15' map, since as the number of stars used to predict the extinction value for a given pixel increases the error on the map is reduced. In addition to the reconstruction variance derived from Equation \ref{eqn:recon_sigma}, we also report a reconstruction variance calibrated with respect to the uncertainty-normalized deviation between the map at FWHM=15' and the value of SFD at pixels at $b>60^\circ$. We do so by adding in quadrature a $\sigma_{sys}=0.01$ to the predicted reconstruction sigma. At this value of $\sigma_{sys}$, the distribution of $(Map@FWHM=15' - SFD)/{\sqrt{Var@FWHM=15' + \sigma_{sys}^2}}$ has a standard deviation of 1.03. If the values of the noise map were truly randomly distributed about the corresponding SFD values, we would have expected a standard deviation of 1. We only chose to set this additive error using higher latitudes where we expect the signal to be reasonably low. One would also not in general expect the uncertainty-normalized bias to have a spread of 1 across the full sky and particularly at lower latitudes with higher extinction and more complex dust morphology where the map's resolution would matter more. We examine the bias of pixels in the map, and the uncertainty-normalized bias ($(Map - SFD)/\sigma(Map)$) of pixels in the map as a function of the extinction values in each pixel in SFD for the overall footprint, in Figure \ref{fig:hoggplot}. The 2D histogram is normalized for each SFD extinction bin, and the values along the x axis for SFD span the $2^{nd}$ to the $98^{th}$ percentile of SFD values in our footprint. The values are most consistent at low values of SFD ($<0.2$mag), which accounts for the majority of the pixels in the footprint we cover. At higher extinction values, all three stellar-reddening maps systematically underpredict extinction relative to SFD. We also note that while such comparisons are useful to gauge consistency across maps, SFD is not a `ground truth' in this comparison. Rather, SFD is an emission-based map while the stellar-reddening based maps measure the effect of extinction — which is the effect that we actually want to map. Furthermore, \cite{yasuda2007spatial} examined SDSS galaxy number counts at lower galactic latitudes, while \cite{arce1999measuring} compared SFD extinction predictions to extinctions derived from the Taurus dark cloud complex using background stars’ color excesses and counts and ISSA 60 and 100 micron images. Both works found that SFD over predicts extinction at $E_{B-V} > 0.15$mag. Other work has also found deviations in SFD: e.g: \cite{peek2010correction} identified corrections to SFD by using quiescent galaxies as standard crayons (objects whose color is known) and found that SFD is underpredicted in regions at lower latitudes and $E_{B-V}<0.15$mag.

\begin{figure}
\centering
    \includegraphics[keepaspectratio=true, width=\linewidth,height=200pt]{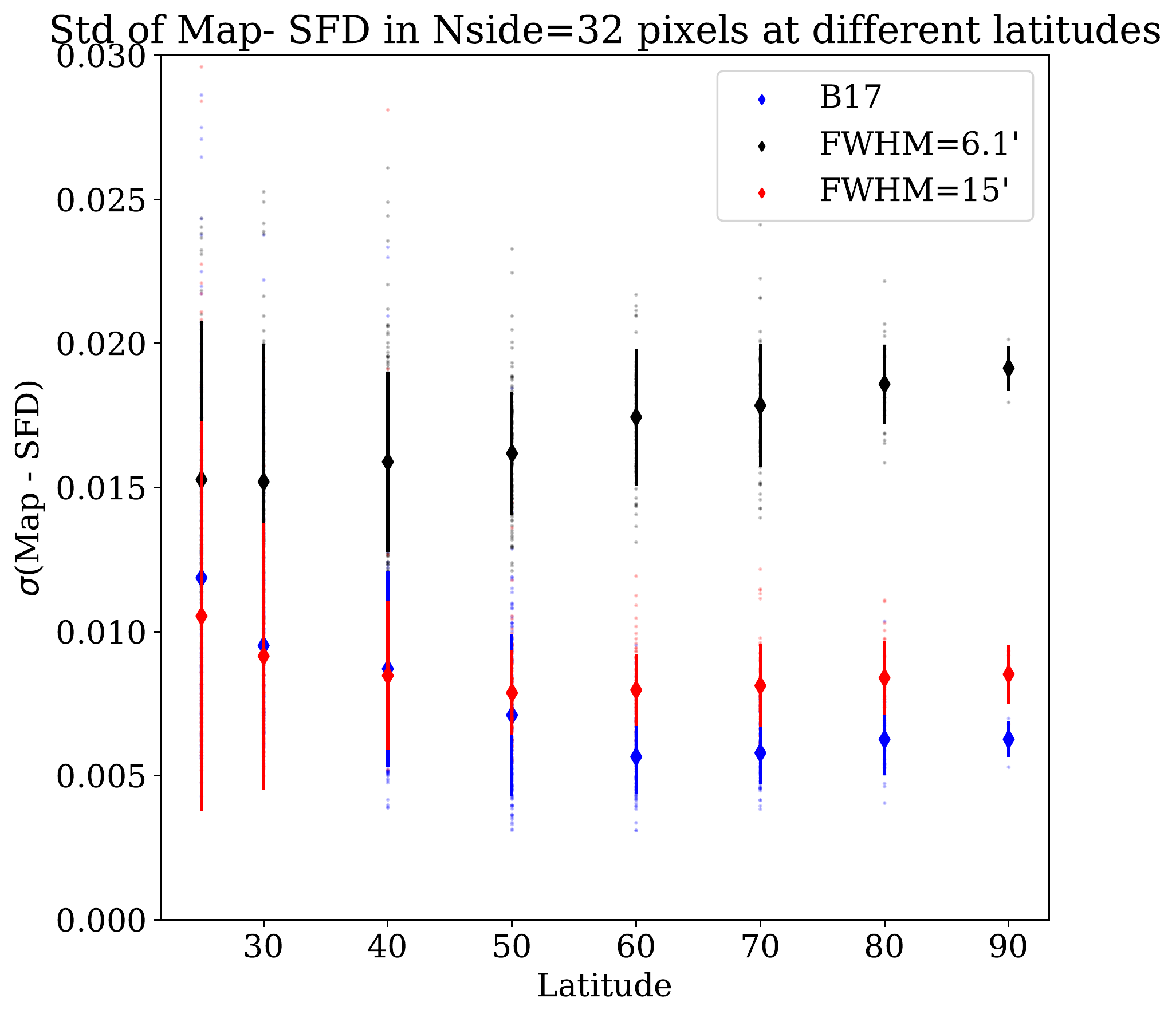} 
\caption{Standard deviations of the offsets of the map relative to SFD for all testbeds. Each datapoint is the standard deviation over the difference between the input map's values and SFD's values in \textsc{Nside}=2048 pixels contained within a single \ns{32} pixel.
\label{fig:latwise_offsets}}
\end{figure}

\begin{figure*}
\centering
    \includegraphics[keepaspectratio=true, width=.99\linewidth]{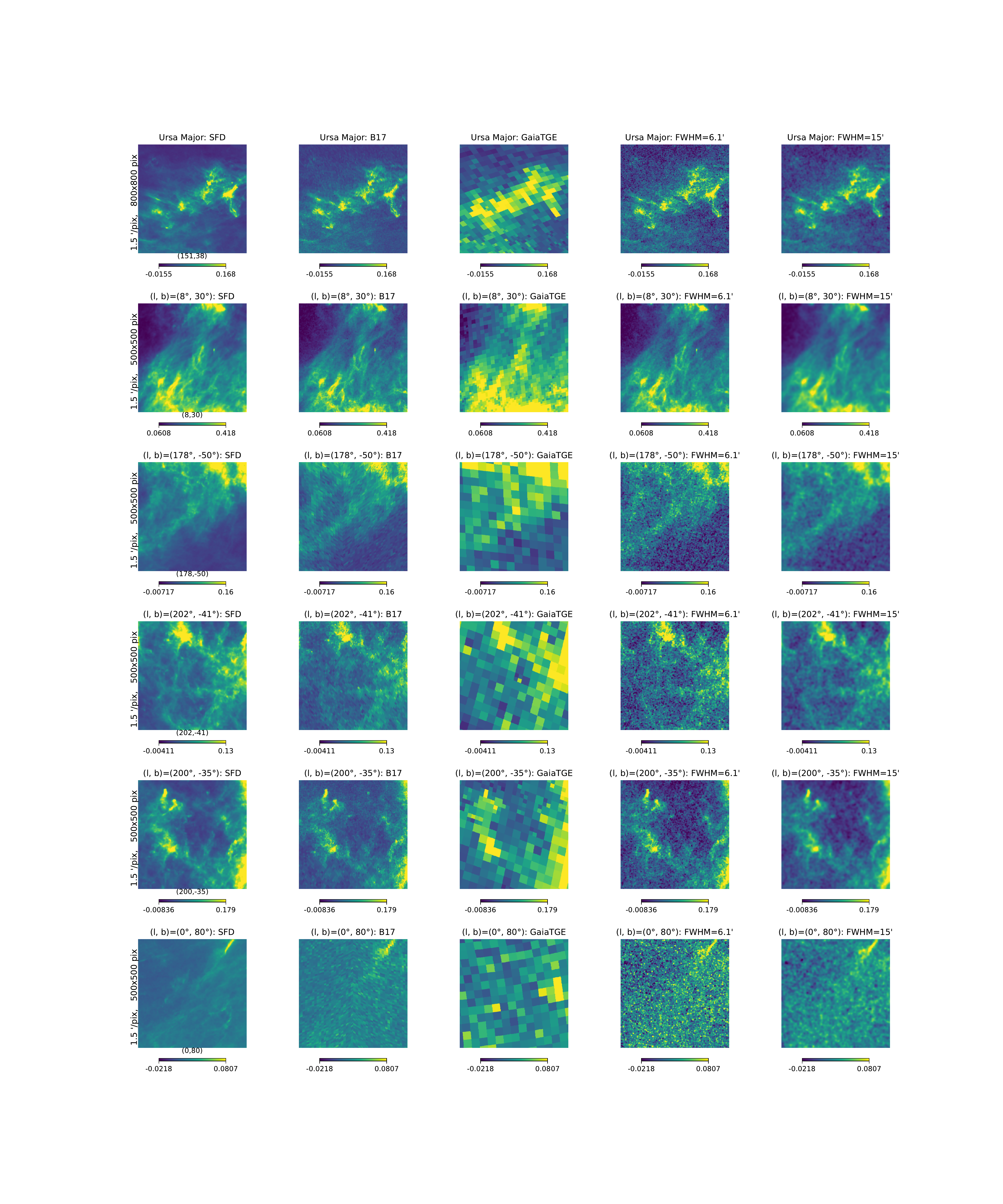}
\caption{Comparison of SFD, \bay{17}, the Gaia TGE map and the reconstructions at FWHM=$6.1'$ and FWHM=$15'$ over different patches of the sky, with the colorscale set by the 1st and 99th percentile of the pixel values of the FWHM=$6.1'$ map for the area under focus.
\label{fig:maps_panel}}
\end{figure*}

\textit{Latitude: } We examine how consistent our maps are with SFD, in \textsc{Nside}=32 pixels at different latitudes in the Galactic Northern Hemisphere. We query all \ns{2048} pixels lying within \ns{32} pixels closest to a given latitude for the following latitudes: $b = [25^{\circ}, 30^{\circ}, 40^{\circ}, 50^{\circ}, 60^{\circ}, 70^{\circ}, 80^{\circ}, 90^{\circ}]$. For each of the \ns{32} pixels in each latitude band, we examine the standard deviation of the difference of the map's extinctions and SFD extinctions. The mean and standard deviation of these values is visualized in Figure \ref{fig:latwise_offsets}. As we move from lower to higher latitudes, with lower dust and relatively fewer stars, the FWHM=$15'$ map is less noisy than the one with FWHM=$6.1'$. In this limit, the less smoothed map is approximately 2.5 times noisier than the map with more smoothing, as would be expected from the ratio of the smoothing scales.

\section{Discussion}\label{sec:discussion}
Contamination of Galactic extinction maps by large scale structure could potentially impact a range of cosmological measurements. \cite{chiang2019extragalactic} demonstrate how galactic dust map cross-correlations with LSS could bias lensing-induced correlations between foreground lenses and background objects and parameter constraints derived from luminosity distances. In the latter case, the bias is still mild, and at the 0.5\% level for $\Omega_m$ and $w$.
 
\cite{kitanidis2020imaging} found that the density of DESI Emission Line Galaxy (ELG) samples and Quasi-Stellar Object (QSO) samples decreases and increases respectively with extinction at a level of nearly 10\% in some extinction bins. In addition to the correlations we focus on in this paper, variation in Galactic extinction and stellar density can also affect the apparent distribution of extragalactic objects and add power on larger scales that are particularly important for constraints on the primordial non-Gaussianity parameter $f_{NL}$ \citep{ross2013clustering, rezaie2021primordial}. Cross correlations, such as those between CMB lensing maps and LSS tracers such as galaxies could also potentially be affected since Galactic dust serves as a common foreground for the CMB, in terms of its emission and for the magnitudes of galaxy catalog objects, in terms of its extinction. \cite{chen2022cosmological} found that the galaxy-CMB lensing cross power spectrum can be biased up to a few percent by extinction even when the bias on the galaxy autopowerspectrum is at the sub percent level. 

The maps we release could serve as a systematics cross-check in several ways. It would be of interest, for example, to examine whether different choices of maps, including the maps we construct, affect the magnitude and direction of these trends.

This work can be built upon in several ways. One of the key limitations of stellar-reddening based maps is their higher noise. In terms of the statistical reconstruction, better regularization could help reduce noise and enable us to derive better posterior uncertainties. The advent of data from upcoming surveys such as LSST would increase the number of stars observed by an order of magnitude and be able to observe much fainter stars ($r\sim27$) \citep{ivezic2019lsst}. LSST data might also render other novel and complimentary ways of constraining galactic dust extinction possible. \cite{bravo2021simultaneous} explored the possibility of simultaneously deriving Galactic dust extinction maps at resolutions of $7'$ and $1^{\circ}$ as well as large scale structure overdensities with a simulated LSST galaxy catalog, a task hitherto considered impossible at resolutions comparable to that of existing emission or extinction-based dust maps because of limitations in depth or sky coverage. With their approach, correlated noise, including correlations between LSS and extinction, become dominant at resolutions of $7'$. They further outline a Bayesian scheme to combine existing dust maps as a prior and galaxy properties as a likelihood to derive a joint posterior distribution over LSS and Galactic extinction. It would be interesting to examine the possibility of combining maps with lower correlations with extragalactic structure such as the ones we sought to make in this paper, with constraints from galaxy properties, to both combine the constraining power of stars and galaxies on the dust distribution at higher latitudes, and minimize correlations with LSS. Better star-galaxy-QSO separation approaches constitute the common underpinning for both the approach we take and the one in \cite{bravo2021simultaneous}, a task that would become more challenging as we go to fainter magnitudes. 

While we were preparing this draft, \cite{delchambre2022gaia} released a total galactic extinction map using Gaia DR3 sources classified as stars from the Discrete Source Classifier (DSC) module. Our work differs from that in several regards: the information used to exclude extragalactic objects, reconstruction choices -- their maps have varying HEALPix resolution, ranging between HEALPix levels 6 and 9, the portion of the sky covered -- they cover the sky at $|b|>5^\circ$ and the stellar inference framework used -- they use \cite{andrae2022gaia} to derive extinctions to each source. Figure \ref{fig:maps_panel} plots SFD, \bay{17}, the Gaia total galactic extinction map, and our maps at FWHM=6.1' and 15' over six different patches of the sky. To derive extinction estimates for the Gaia TGE map we divide the map, which reports extinction at 541.4 nm assuming the Fitzpatrick extinction law with $R_V=3.1$. Having multiple extinction maps allows us to test for sensitivity to differences and assumptions of any single stellar inference framework or source selection scheme.

\section{Conclusion}
Minimizing systematics in cosmological analyses is important to be able to harness the full power of datasets from DESI and LSST in the near future. To this end, we focused on examining the question of whether removing suspected extragalactic objects from stellar catalogs reduces correlations with large scale structure, and release two resulting Galactic extinction maps. We cross-match to SDSS DR17 data and learn the boundaries of regions populated by stars in PS1 \textit{riz} - WISE W1 and SDSS \textit{ugr} space. We additionally have cuts on Gaia EDR3 parallax and the reduced chi-squared statistic of the objects in the catalog with stellar models. Since we choose to reconstruct our maps in two dimensions, and wish to use only stars lying behind the sheet of integrated Galactic extinction over the sky, we add distance cuts to select only stars that we expect to lie behind dust in the Galaxy.
This set of stellar selections reduces QSO contamination from 15\% to $\sim0.1$\% on a test set of spectroscopically matched objects in the Southern Galactic Hemisphere. We then apply these cuts on the set of stellar catalog objects and construct two dust maps at a Full-Width Half-Maximum of 6.1’ and 15’ for the intersection of the PS1 footprint and the region with $|b|>20^\circ$. We evaluate the angular cross-correlation of these maps and for existing dust maps, using the clustering-redshift based technique, and find that our map has lower correlations than most existing stellar-reddening and emission-based maps. 

\section{Code and Data Availability}
We make the code used for this work available on Github at \href{https://github.com/nmudur/HighLatMaps/}{\textbf{HighLatMaps}}. The maps used for this work can be found here at this Zenodo link: \url{https://doi.org/10.5281/zenodo.7411344}

\section{Acknowledgements}
We are especially grateful to Gregory M. Green and Catherine Zucker for help with running the Bayestar stellar inference code and useful discussions. We thank Tanveer Karim, Andrew K. Saydjari, Joshua S. Speagle, Justina R. Yang and Ioana Zelko, for helpful discussions and inputs. We thank Yi-Kuan Chiang, Arjun Dey, Daniel Eisenstein, Ashley Ross, and David Schlegel for insightful conversations on this work. This work was supported by the National Science Foundation under Cooperative Agreement PHY2019786 (The NSF AI Institute for Artificial Intelligence and Fundamental Interactions).  D.P.F. acknowledges support by NASA ADAP grant 80NSSC21K0634 “Knitting Together the Milky Way: An Integrated Model of the Galaxy’s Stars, Gas, and Dust.” C.F.P. acknowledges the support of NIH R01.

The Pan-STARRS1 Surveys (PS1) and the PS1 public science archive have been made possible through contributions by the Institute for Astronomy, the University of Hawaii, the Pan-STARRS Project Office, the Max-Planck Society and its participating institutes, the Max Planck Institute for Astronomy, Heidelberg and the Max Planck Institute for Extraterrestrial Physics, Garching, The Johns Hopkins University, Durham University, the University of Edinburgh, the Queen's University Belfast, the Harvard-Smithsonian Center for Astrophysics, the Las Cumbres Observatory Global Telescope Network Incorporated, the National Central University of Taiwan, the Space Telescope Science Institute, the National Aeronautics and Space Administration under Grant No. NNX08AR22G issued through the Planetary Science Division of the NASA Science Mission Directorate, the National Science Foundation Grant No. AST-1238877, the University of Maryland, Eotvos Lorand University (ELTE), the Los Alamos National Laboratory, and the Gordon and Betty Moore Foundation. 

This publication makes use of data products from the Two Micron All Sky Survey, which is a joint project of the University of Massachusetts and the Infrared Processing and Analysis Center/California Institute of Technology, funded by the National Aeronautics and Space Administration and the National Science Foundation.

This work has made use of data from the European Space Agency (ESA) mission
{\it Gaia} (\url{https://www.cosmos.esa.int/gaia}), processed by the {\it Gaia}
Data Processing and Analysis Consortium (DPAC,
\url{https://www.cosmos.esa.int/web/gaia/dpac/consortium}). Funding for the DPAC
has been provided by national institutions, in particular the institutions
participating in the {\it Gaia} Multilateral Agreement.

Funding for the Sloan Digital Sky 
Survey IV has been provided by the 
Alfred P. Sloan Foundation, the U.S. 
Department of Energy Office of 
Science, and the Participating 
Institutions. 

SDSS-IV acknowledges support and 
resources from the Center for High 
Performance Computing  at the 
University of Utah. The SDSS 
website is www.sdss.org.

SDSS-IV is managed by the 
Astrophysical Research Consortium 
for the Participating Institutions 
of the SDSS Collaboration including 
the Brazilian Participation Group, 
the Carnegie Institution for Science, 
Carnegie Mellon University, Center for 
Astrophysics | Harvard \& 
Smithsonian, the Chilean Participation 
Group, the French Participation Group, 
Instituto de Astrof\'isica de 
Canarias, The Johns Hopkins 
University, Kavli Institute for the 
Physics and Mathematics of the 
Universe (IPMU) / University of 
Tokyo, the Korean Participation Group, 
Lawrence Berkeley National Laboratory, 
Leibniz Institut f\"ur Astrophysik 
Potsdam (AIP),  Max-Planck-Institut 
f\"ur Astronomie (MPIA Heidelberg), 
Max-Planck-Institut f\"ur 
Astrophysik (MPA Garching), 
Max-Planck-Institut f\"ur 
Extraterrestrische Physik (MPE), 
National Astronomical Observatories of 
China, New Mexico State University, 
New York University, University of 
Notre Dame, Observat\'ario 
Nacional / MCTI, The Ohio State 
University, Pennsylvania State 
University, Shanghai 
Astronomical Observatory, United 
Kingdom Participation Group, 
Universidad Nacional Aut\'onoma 
de M\'exico, University of Arizona, 
University of Colorado Boulder, 
University of Oxford, University of 
Portsmouth, University of Utah, 
University of Virginia, University 
of Washington, University of 
Wisconsin, Vanderbilt University, 
and Yale University.

This publication makes use of data products from the Wide-field Infrared Survey Explorer, which is a joint project of the University of California, Los Angeles, and the Jet Propulsion Laboratory/California Institute of Technology, funded by the National Aeronautics and Space Administration.

\textit{Software:} GNU Parallel \citep{tange_ole_2018_1146014}, \textsc{Numpy} \citep{harris2020array}, \textsc{Astropy} \citep{astropy:2018}, \textsc{HealP}ix \citep{Zonca2019}, \textsc{Scikit-Learn} \citep{scikit-learn}, \textsc{Pandas} \citep{reback2020pandas}, \textsc{Matplotlib} \citep{Hunter:2007}, \textsc{Dustmaps} \citep{2018JOSS....3..695M}

\onecolumngrid
\appendix
\section{Details of Stellar Selections}
\subsection{Preliminary Selections}
The cuts on each object require all of the following:
\begin{itemize}
    \item \texttt{nmag\_ok} $>0$ in at least 2 PS1 bands and the sum of \texttt{nmag\_ok} over all 5 PS1 bands $\geq 4$
    \item $m^{PS1}_{PSF} - m^{PS1}_{Aperture} < 0.1$ in at least 2 bands
    \item `good detections' in at least 4 out of 8 bands
    \item `good detections' in at least 2 PS1 bands
    \item not extended in any of the 2MASS bands, using the \texttt{ext\_key} flag
\end{itemize}

A `good detection' for each PS1 band passes the following quality cuts: \texttt{nmag\_ok}$>0$ AND the magnitude is fainter than the PS1 saturation limit AND the error is less than 0.2. A `good detection' in each 2MASS band satisfies C1 AND C2, where C1 and C2 are conditions defined for each band as follows:
\begin{itemize}
    \item C1 $\leftarrow$ (\texttt{ph\_qual}=`A') OR (\texttt{rd\_flg}=1) OR (\texttt{rd\_flg}=3) AND (\texttt{cc\_flg}=0)
    \item C2 $\leftarrow$ (\texttt{use\_src}=1) AND (\texttt{gal\_contam}=0)
\end{itemize}
The \texttt{ph\_qual} flag for each band is a measure of the quality of photometry for that band, on the basis of \texttt{rd\_flg}, scan signal to noise ratios and measurement uncertainties. The definitions of these flags and criteria can be found in \cite{cutri2006explanatory}. The choice of PS1 and 2MASS cuts above is the same as that followed in \cite{Green2019}.

The criterion for using Gaia EDR3 parallaxes is as follows:
\begin{itemize}
    \item the Renormalized Unit Weight Error (RUWE) must be less than 1.4
    \item the prediction of the astrometric fidelity classifier must exceed 0.5
\end{itemize}

\subsection{SVM based cuts}
To generate a spectroscopically matched subset of objects, we query all stars at $b>50^\circ$ from the catalog obtained after the preliminary set of selections (Section 3) and cross match to the SDSS DR17 specobj catalog. This gives us a much smaller subset of objects, all of which have spectroscopic labels (\texttt{sdss\_dr17\_specobj.CLASS}). The class of an object is either `STAR', `GALAXY' or `QSO' -- for both models, an object with a class of `STAR' is assigned a label of 1, and 0 otherwise. We perform the following quality cuts to identify objects with reliable spectra -- \texttt{rchi2}$<2$, \texttt{chi68p}$<2$, and \texttt{sn\char`_median\char`_all}$>10$.

To train the WISE SVM cut, we further identify objects with reliable magnitude measurements for the relevant features by requiring the magnitude errors in the PS1 bands $r, i$ and $z$ and WISE W1 band to be less than 0.2 and \texttt{w1rchi2}$<2$, yielding 158154 objects. We then generate a balanced training set consisting of an equal number of objects with labels 0 and 1. We then train an SVM with a linear kernel, that takes two features as its input: $m_r - m_i$ and $m_z - W1$, using the \textsc{Scikit-Learn} package to learn the boundary in the left panel of Figure \ref{fig:models}. When applied to an arbitrary object in the full catalog, objects with values of z-W1 lying above the SVM boundary were eliminated. The cut is implemented such that objects without a detection in W1 (\texttt{allwise.w1mpro}=0) are assigned a `limiting' magnitude $W1_{lim}$=17.44, which corresponds to the 96th percentile of objects with reliable spectra and detections in W1. The rationale behind this is that an object without a detection in W1 would be fainter than $W1_{lim}$, or that if it had been detected, $W1>W1_{lim}$. Thus its $z-W1>z-W1_{lim}$ and consequently if $z-W1_{lim}$ lies above the SVM boundary, so would $z-W1$.

To train the SDSS SVM based cut, we take the same set of spectroscopically labelled objects as above and perform the same cuts to select objects with reliable spectra. We further require magnitude errors in the $u, g, $ and $r$ bands from SDSS DR14 starsweep to be less than 0.2 and filter out objects with invalid (NaN) magnitudes or with magnitudes of exactly 22.5 in $u, g, $ or $r$, yielding 203492 objects. As above, we balance the training set to consist of an equal number of objects with label 0 or 1. We then train an SVM based classifier with a radial basis function kernel that takes two features as it's input: $m_u - m_g$ and $m_g - m_r$. 

\begin{deluxetable*}{cccc}
\tablenum{3}
\tablecaption{Distribution of objects from the spectroscopically matched sample in the Northern Galactic Hemisphere in different classes as stellar selection cuts are incrementally applied. The number reports the number of objects in each class. The figure in brackets reports the fraction of objects belonging to a certain category after a particular choice of cuts is applied $\frac{n^{After Cuts}_{Class}}{n^{After Cuts}_{Total}}$. Each row in the table also includes all the cuts in the rows preceding it. (The percentages in each row sum to 100\%). \label{tab:spec_classes}}
\tablewidth{0pt}
\tablehead{
\colhead{Cuts} & \colhead{Stars: No./ Proportion (\%)} & \colhead{QSOs: No./ Proportion (\%)} & \colhead{Galaxies: No./ Proportion (\%)} 
}
\startdata
No Cuts Applied & 175597 (83.33) & 34817 (16.52) & 299 (0.14) \\
+$d_{50}sin(b) > 400pc, \sigma_{\mu}<1.5$ & 162960 (83.53) & 31839 (16.32) & 280 (0.14) \\
+WISE nondetection/(r-i, z-W1) cut & 158629 (98.91) & 1561 (0.97) & 190 (0.12) \\
+ eDR3 parallax detection & 158311 (98.94) & 1561 (0.98) & 137 (0.08) \\
+SDSS (u-g, g-r) cut & 155897 (99.80) & 203 (0.13) & 109 (0.07) \\
+$\chi^2_r<3$ & 138797 (99.93) & 46 (0.03) & 51 (0.04) \\
\enddata
\end{deluxetable*}

\begin{deluxetable*}{cccc}
\tablenum{4}
\tablecaption{Ablation Tests on the spectroscopically matched sample in the Northern Galactic Hemisphere: All cuts in the Secondary Stellar Selection EXCEPT a given cut are applied.  \label{tab:ablation}}
\tablewidth{0pt}
\tablehead{
\colhead{Excluded Cut} & \colhead{Stars: Number} & \colhead{QSOs: Number} & \colhead{Galaxies: Number } 
}
\startdata
No cuts excluded & 138797  & 46 & 51 \\
\hline
-WISE nondetection/(r-i, z-W1) cut & 139957 & 103 & 57 \\
-eDR3 parallax detection & 139059 & 46 & 62 \\
-SDSS (u-g, g-r) cut & 139725 & 232 & 61 \\
-$\chi^2_r<3$ & 155897 & 203 & 109 \\
\enddata
\end{deluxetable*}

\begin{figure*}[!h]
\centering
    \includegraphics[keepaspectratio=true, width=.49\linewidth,height=800pt]{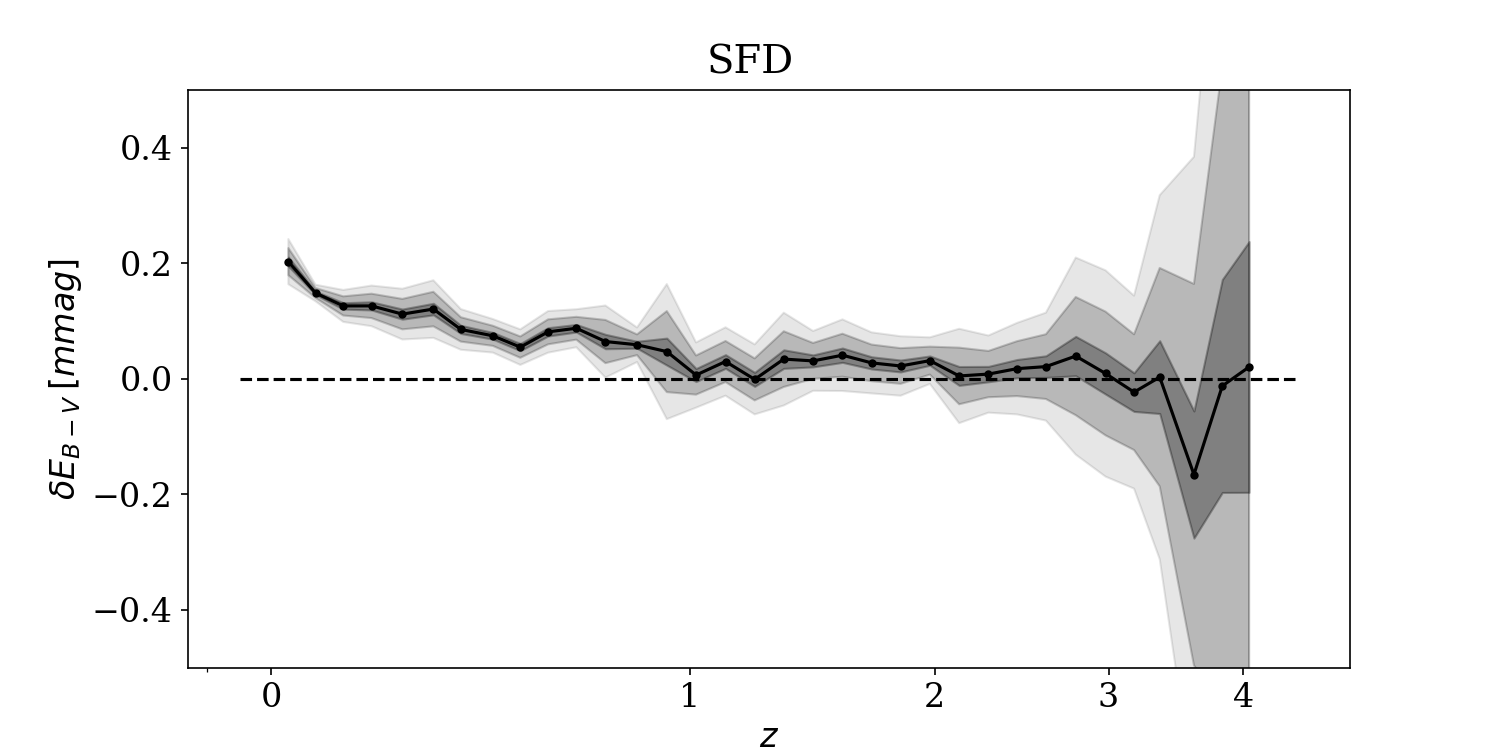}
    \includegraphics[keepaspectratio=true, width=.49\linewidth,height=800pt]{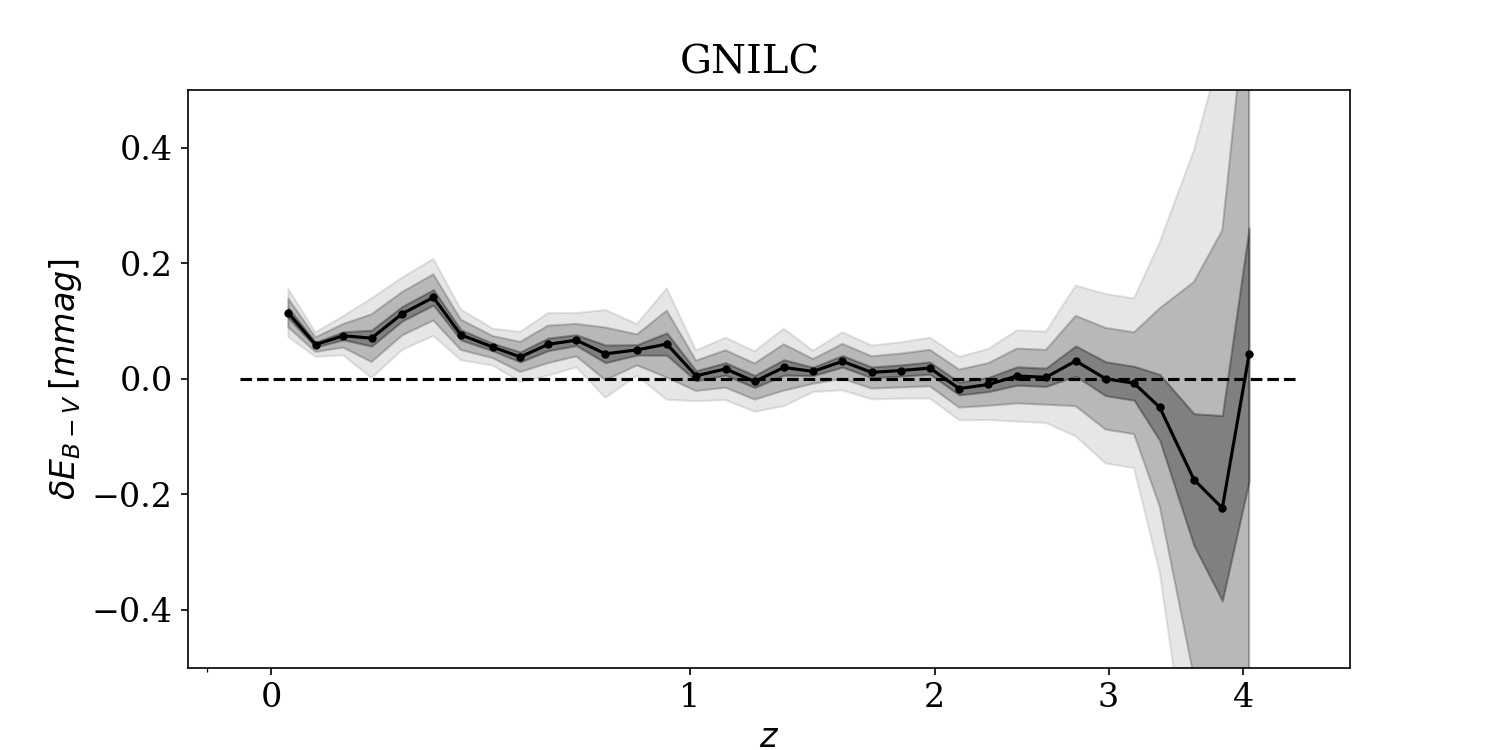}    
    \includegraphics[keepaspectratio=true, width=.49\linewidth,height=800pt]{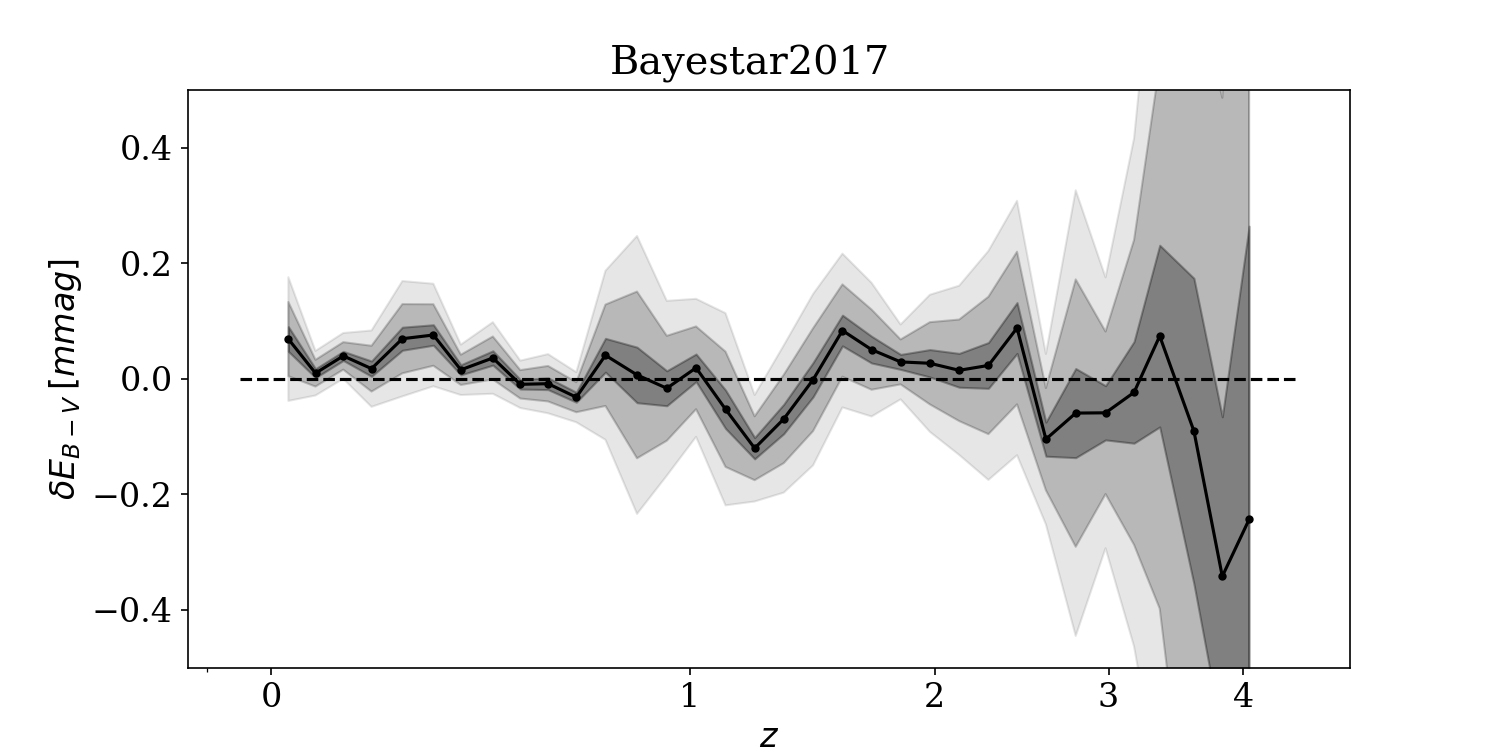}
    \includegraphics[keepaspectratio=true, width=.49\linewidth,height=800pt]{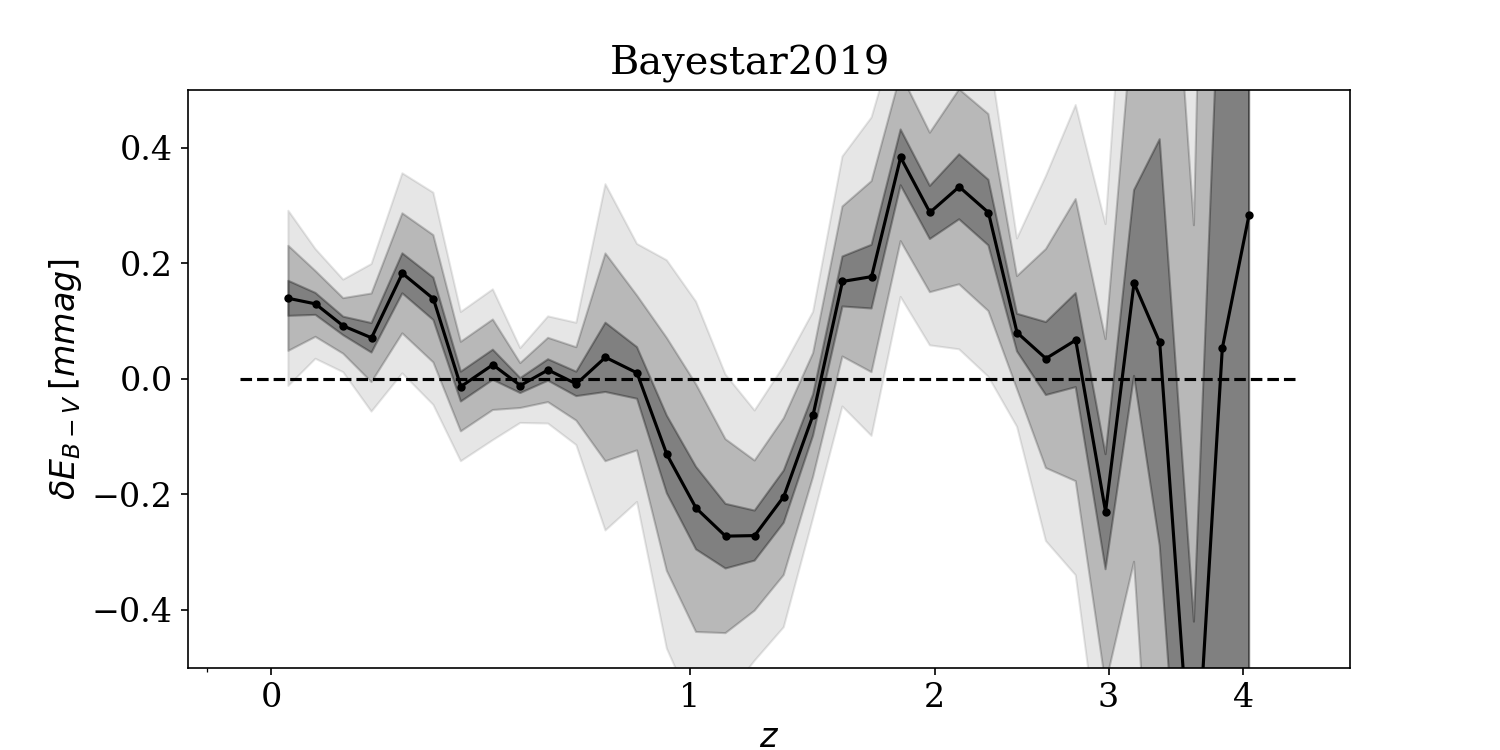}
    
    \includegraphics[keepaspectratio=true, width=.49\linewidth,height=800pt]{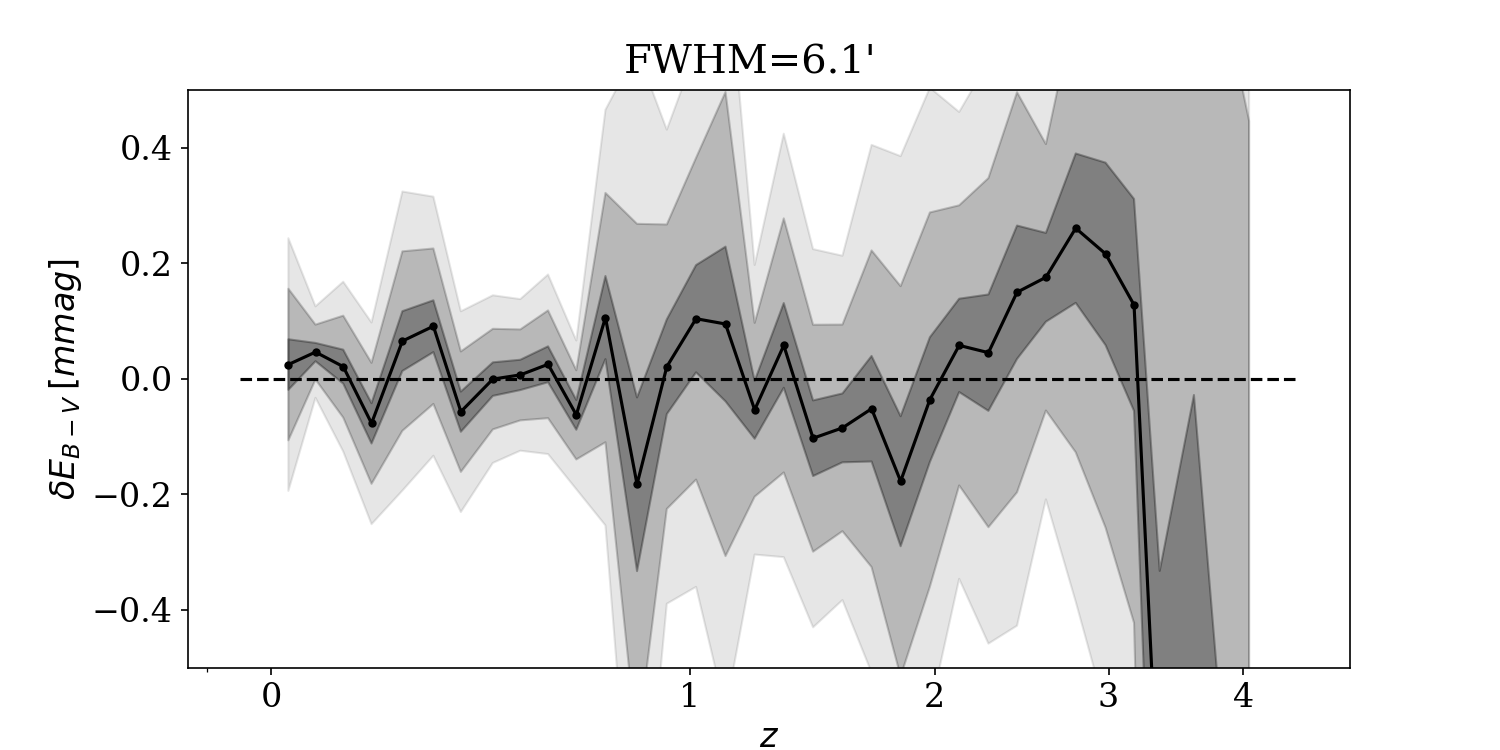}
    \includegraphics[keepaspectratio=true, width=.49\linewidth,height=800pt]{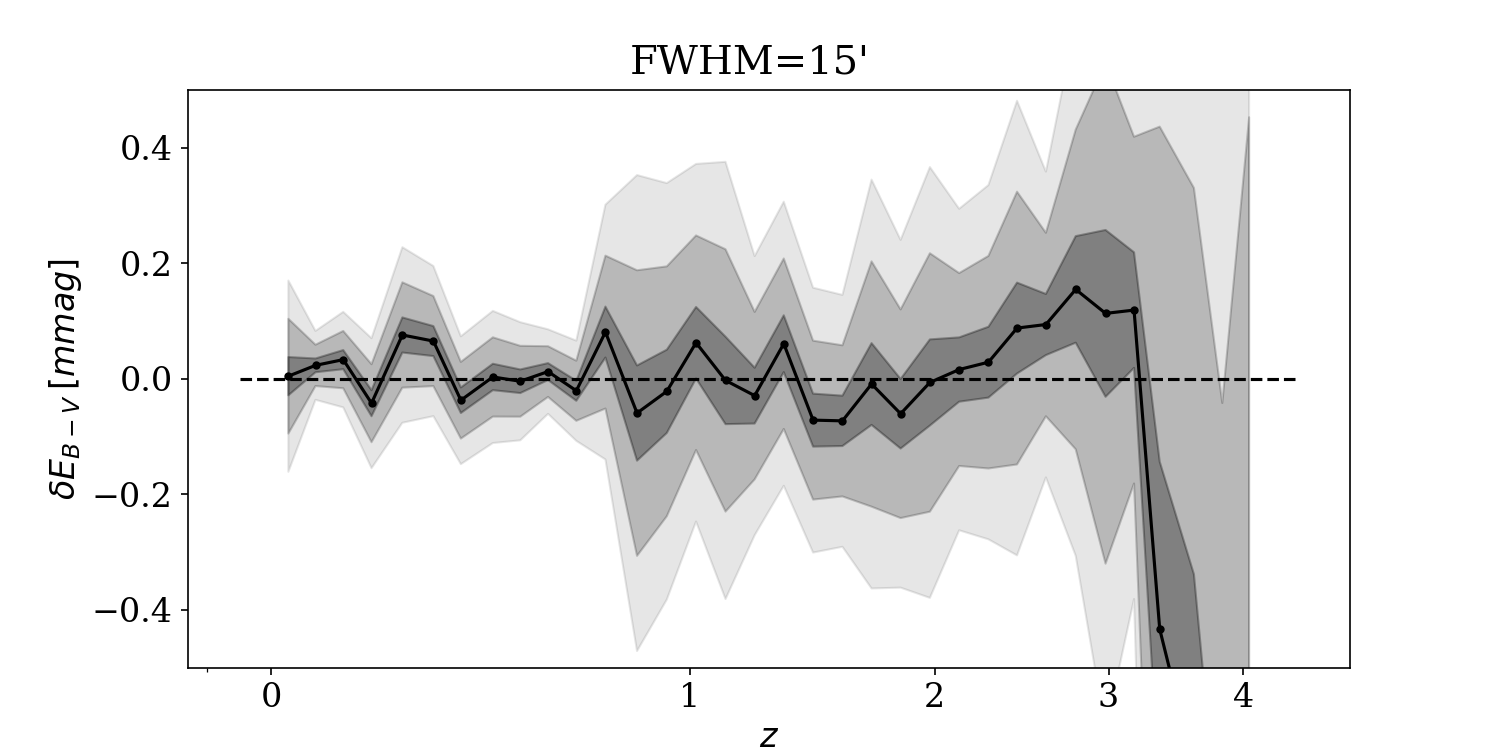}
\caption{Excess extinction in maps at the positions of reference extragalactic objects in redshift bins spanning z=0-4 for existing emission-based maps (top panel: SFD and the Planck 2016 (GNILC)) dust emission map, existing stellar-reddening based maps (middle panel: \bay{17} and \bay{19}) and our reconstructions with FWHMs of 6.1’ and 15’ (last panel). The sigma contours are obtained by bootstrapping over the reference sample objects at each redshift.
\label{fig:acc_main_bootstrapped}}
\end{figure*}

\onecolumngrid
\subsection{Deriving the angular cross correlation estimator}\label{sec:acc_deriv}
In this section, we describe how the angular cross correlation is computed, largely following the prescription in \cite{chiang2019extragalactic}. $\phi$ indicates a position (pixel) on the sky and $E(\phi)$ an input dust map. $\phi + \theta$ denotes all coordinates lying at an angular separation of $\theta$ from the pixel $\phi$. $\delta_r (\phi, z)$ denotes the fractional overdensity of all reference extragalactic objects in redshift bin $z$, (denoted by $\mathcal{R}(z)$), lying in the pixel $\phi$. Thus, $\delta_{r} (\phi + \theta, z)$ is the fractional overdensity field of reference objects belonging to $\mathcal{R}(z)$
evaluated in the ring with its center at  $\phi$  and radius $\theta$. $\langle\dele^{EG}(\theta, z)\rangle_r$ denotes the average (over all pixels in the masked region) excess extinction at an angular separation of $\theta$ around reference extragalactic objects at redshift $z$. $\langle\Delta E (z)\rangle$ is the integral of $\langle\dele^{EG}(\theta, z)\rangle_r$ over all angular bins from 0 to $\theta_{max}$ divided by $\theta_{max}$ and is the $\delta E_{B-V}$ quantity plotted on the y axis, as a function of $z$ in Figures \ref{fig:null}, \ref{fig:acc_main}, \ref{fig:ablationacc}, and \ref{fig:acc_main_bootstrapped}.
\begin{align}
& \dele (\phi) \leftarrow E(\phi) - {\rm Smooth}(E(\phi))_{\sigma_{GAL}} \label{eqn:delE}\\
&  \dele (\phi) \leftarrow  \dele (\phi) - \langle\dele (\phi)\rangle \\
& \langle\dele^{EG}(\theta, z)\rangle_r = \langle\delta_r (\phi, z) \dele (\phi + \theta)\rangle_\phi \label{eqn:acc_ztheta}\\
& \langle\Delta E (z)\rangle = \frac{1}{\theta_{max}}\int_0^{\theta_{max}} \langle\dele^{EG} (\theta, z)\rangle_r d\theta = \vec{W}_z^T \vec{\dele} \label{eqn:acc_z}
\end{align}

The cross-correlation weight matrix for a given redshift $z$ is stored as  $W_z(\phi)$ where $\phi$ is the set of pixels at \textsc{Nside=2048} over which the cross-correlation signal is calculated. 

\begin{align}
& \langle\Delta E (z)\rangle = \frac{1}{\theta_{max}}\int_0^{\theta_{max}} \langle\dele^{EG} (\theta, z)\rangle_r d\theta = \frac{1}{\theta_{max}} \int_0^{\theta_{max}} d\theta \langle\delta_r (\phi + \theta, z) \dele (\phi)\rangle_\phi \nonumber\\
&= \frac{1}{\theta_{max}} \sum_{\theta_i=0}^{\theta_{max}} \Delta\theta_i \sum_\phi \frac{1}{N(z) N_{pixels}} \sum\limits_{r \in \mathcal{R}(z)}  \delta^K(r, \phi+\theta_i) \dele (\phi) = \vec{W}_z^T \vec{\dele}  \label{eqn:acc_intg} \\
& \text{ where, } W_z(\phi) = \frac{\sum\limits_{r \in \mathcal{R}(z)} \sum_{\theta_i=0}^{\theta_{max}}  \Delta\theta_i   \delta^K(r, \phi+\theta_i)}{\theta_{max}N(z)N_{pixels}} \text{ and } N(z) = |\mathcal{R}(z)| \nonumber
\end{align}

$\delta^K(r, \phi+\theta_i) = 1$ if reference object \textit{r} lies in the ring with its center at  $\phi$ and radius $\theta_i$, 0 otherwise.
$N_{pixels}$ is the number of pixels in the mask on the map over which the correlation is evaluated and $\Delta\theta_i$ is the integration coefficient corresponding to the angular bin $\theta_i$.

\bibliography{ref}
\bibliographystyle{aasjournal}



\end{document}